%% file: wheen_paper.tex
\documentclass[a4paper]{article}
\usepackage[utf8]{inputenc}
\usepackage{pgfplots}
\pgfplotsset{compat=newest,width=5cm}
\usepackage[a4paper,margin=25mm]{geometry}    % page layout
\usepackage{setspace} 
\usepackage{amsfonts,amssymb,amsmath}         % useful math extensions
\usepackage{textcomp}
\usepackage{graphicx}                         % graphics import
\usepackage{wasysym}
\usepackage{import}
\usepackage{nicefrac}
\usepackage{appendix}
\graphicspath{{figs/}}

\makeatletter
\def\input@path{{figs/}}
\makeatother

\pdfpageattr{/Group << /S /Transparency /I true /CS /DeviceRGB>>}
\usepackage[colorlinks=true, allcolors=black]{hyperref}
\usepackage{hhline}
\usepackage{subcaption}
\usepackage{multirow}
\usepackage[toc]{multitoc}
\newcommand\numberall{\addtocounter{equation}{1}\tag{\theequation}}
\title{\huge\bfseries{\vspace{-75pt}The largely linear response of earth's ice volume to orbital forcing}}
\author{Liam Wheen, Thomas Gernon, Cameron Hall, Jerry Wright, and Oscar Benjamin}
\date{\today}
\begin{document}
\begin{titlepage}
  \maketitle
\begin{abstract} Orbital forcing
plays a key role in pacing the glacial-interglacial cycles. However, the
mechanistic linkages between the orbital parameters — eccentricity, obliquity,
and precession — and global ice volume remain unclear. Here, we investigate the
effect of Earth’s orbitally governed incoming solar radiation (that is,
insolation) on global ice volume over the past 800,000 years. We consider a
simple linear model of ice volume that imposes minimal assumptions about its
dynamics. We find that this model can adequately reproduce the observed ice
volume variations for most of the past 800,000 years, with the notable exception
of Marine Isotope Stage 11. This suggests that, aside from a few extrema, the
ice volume dynamics primarily result from an approximately linear response to
orbital forcing. We substantiate this finding by addressing some of the key
criticisms of the orbitally forced hypothesis. In particular, we show that
eccentricity can significantly vary the ocean temperature without the need for
amplification on Earth. We also present a feasible mechanism to explain the
absence of eccentricity's 400,000 year period in the ice volume data. This
requires part of the forcing from eccentricity to be lagged via a
slow-responding mechanism, resulting in a signal that closer approximates the
change in eccentricity. A physical interpretation of our model is proposed,
using bulk ocean and surface temperatures as intermediate mechanisms through
which the orbital parameters affect ice volume. These show reasonable alignment
with their relevant proxy data, though we acknowledge that these variables
likely represent a combination of mechanisms.
\end{abstract}
\begin{spacing}{0.9}
\tableofcontents
\end{spacing}
\end{titlepage}
\pagenumbering{arabic}
\section{Introduction}
\label{sec:introduction}
\begin{figure}
    \centering
    \input{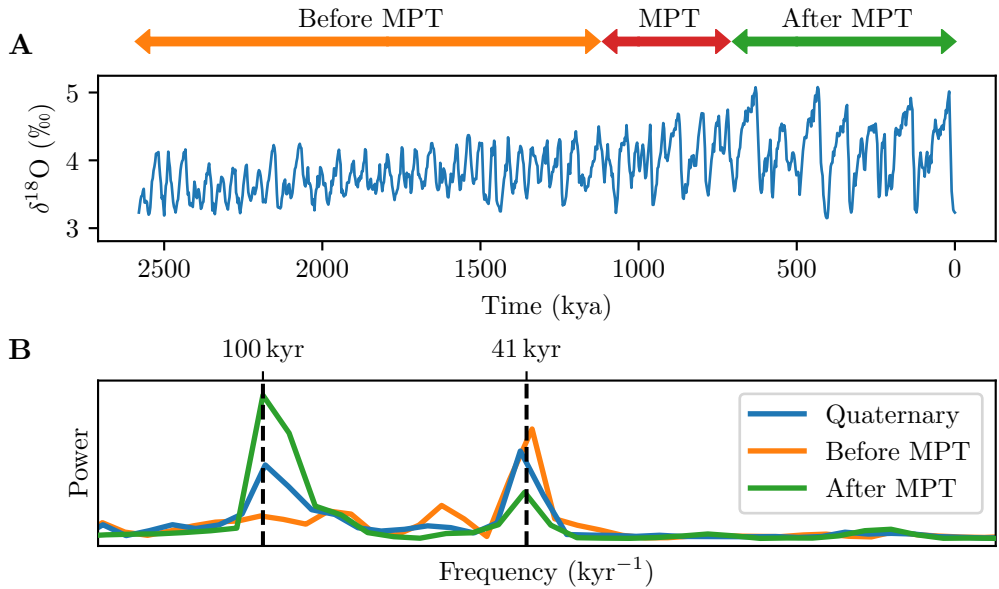}
    \caption{Time series (\textbf{A}) and power spectra (\textbf{B}) for the
      benthic $\delta^{18}$O stack from Lisiecki and Raymo covering the
      Quaternary period~\cite{benthic_data}. The benthic $\delta^{18}$O ratio is
      commonly used as a proxy for global ice volume. The MPT spans
      approximately 550\,kyr and marks a distinct change in dominant period,
      from 41\,kyr to 100\,kyr, as shown in the power spectra.}
  \label{fig:benth_and_power_specs}
\end{figure}

For the past 800 thousand years (kyr), Earth's global ice volume has varied with
a dominant period of approximately 100\,kyr. These oscillations are referred to
as the glacial-interglacial cycles and are demonstrated in
Fig.~\ref{fig:benth_and_power_specs}. This figure also shows the
mid-Pleistocene transition (MPT), which took place from approximately 1250 to
700 thousand years ago (kya)~\cite{mpt_duration}. This marks a significant
change in ice volume dynamics whereby the dominant period shifted from 41\,kyr
to 100\,kyr.

There is a clear link between Earth's orbital parameters and ice volume
dynamics, as demonstrated by the power spectra in Fig.
\ref{fig:orbital_and_benthic_time_series_power_specs}. Each of the prominent
frequencies of ice volume appear to align with those of the orbital
parameters. It is well understood that Earth's orbital configuration varies the
magnitude and distribution of incoming solar radiation, known as insolation.
However, the way in which this impacts Earth's cryosphere is still
unclear. Most hypotheses can be categorised into two schools of thought. The
first attributes the glacial-interglacial cycles to Earth-based mechanisms, with
the orbital parameters entraining the free oscillation. We refer to this as the
\textit{Intrinsic Forcing with Orbital Entrainment} (IFOE) approach. Examples of
this intrinsic forcing are interactions between the ice sheet and
bedrock~\cite{intrinsic_precession_freq_diff}, CO$_2$ variation from geological
sources and sinks~\cite{saltzman_intrinsic,pp04}, and dust impacting Earth's
surface albedo~\cite{intrinsic_dust}. The opposing school of thought attributes
the glacial-interglacial cycles to orbitally governed variations in insolation,
with the potential for feedback mechanisms on Earth to amplify the effect. We
refer to this as the \textit{Orbital Forcing with Potential Amplification}
(OFPA) approach. In this paper, we present a model that provides evidence for
OFPA, as well as attempting to address some of the key limitations of the
approach.
\subsection{Background}
\label{sec:background}
Croll first proposed that the Earth's orbit influenced the
glacial-interglacial cycles through insolation in the late 19th
century~\cite{croll_before_milanko}. However, they incorrectly concluded that
the ice sheets vary asynchronously across the hemispheres. With the advent of more
reliable data, Milankovitch was able to show that the glacial periods occur
simultaneously around the globe, with ice volume variations being more
pronounced in the northern hemisphere~\cite{milankovitch}.

Since then, a number of IFOE and OFPA models have been developed to simulate ice
volume in response to the orbitally governed insolation. A common measure used
to represent this forcing is the average daily insolation on the summer solstice
at 65$^\circ$ north, known as $Q_{65}$. In 1980, Imbrie and Imbrie modelled the
change in ice volume as proportional to $Q_{65}$, supporting the OFPA school of
thought~\cite{imbrie}. Realising that the rate of ice growth is slower than its
recession, they included a conditional time constant that switches to capture
the two rates.

In 2004, Paillard and Parrenin proposed a more complex system that was able to
reproduce the MPT with a sliding parameter~\cite{pp04}. This model also included
$Q_{65}$ as an input and a switching mechanism that depended on the direction of
ice volume change. They also included a variable that represents deep-water
stratification, which depends on ice volume induced changes in salinity. This
feeds into a variable for atmospheric CO$_2$ which, in turn, affects ice volume,
allowing the model to produce unforced oscillations. This makes the model
aligned with IFOE school of thought.

A later paper from Imbrie (2011) focussed on the interplay between the orbital
parameters and how this affects ice volume, again changing the dynamics
depending on the direction in which the ice is changing~\cite{imbrie2011}.
Though this supports the OFPA approach, it can also produce free oscillations,
meaning intrinsic dynamics could be captured by the model. An important
result of this paper is that it was able to recreate the MPT without using a
time dependent parameter. This model produces the closest fit of the three
models discussed so far, but is very sensitive to parameter perturbations.

In the same year, Crucifix developed a Van der Pol style model that produces free
oscillations with a 100\,kyr period, which is entrained by the $Q_{65}$
signal~\cite{crucifix_original}. This model aligns with the IFOE approach,
though the main motivation was to demonstrate that the ice volume dynamics are
easily produced by a carefully tuned model. However, if the model is highly
sensitive to parameter perturbations, its predictions may be unreliable.

A common trait amongst these models is the assumption of a switching mechanism
that depends on whether the ice volume is growing or ablating. This is
reasonable, given the sawtooth nature of the ice volume data in places (Fig.
\ref{fig:benth_and_power_specs}A), however, there is no consensus on what this
mechanism is. The inclusion of a switching mechanism also means that the model
becomes non-linear and highly sensitive to the choice of switching condition.

\subsection{Outline}
\label{sec:outline}
\begin{figure}
    \centering
    \input{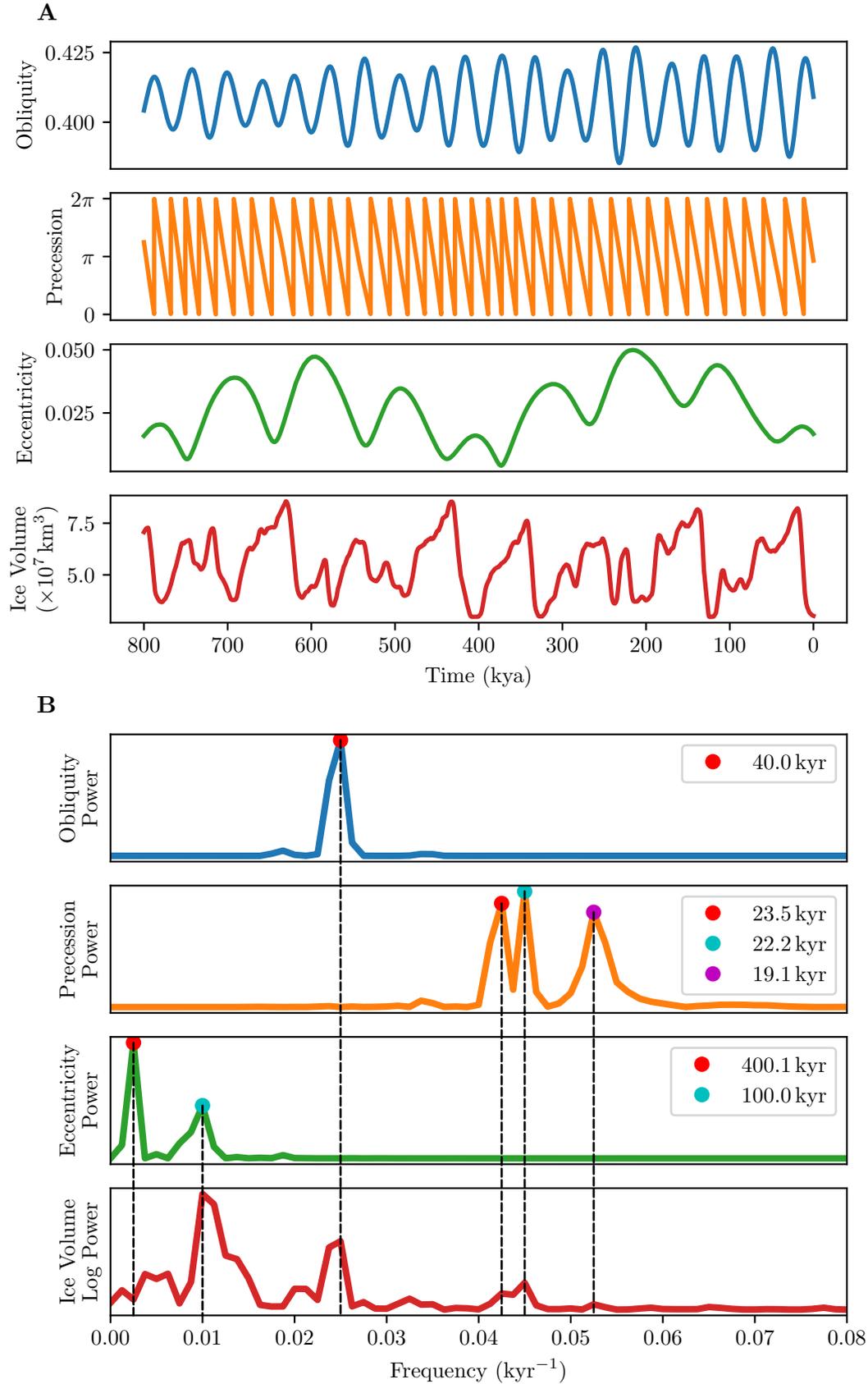}
    \caption{Time series (\textbf{A}) and power spectra (\textbf{B}) for the
      three orbital parameters and ice volume data, the sources for which are
      discussed in Appendices \ref{sec:orbital_data} and \ref{sec:ice_vol}. The
      ice volume power is logarithmically scaled to highlight the smaller peaks
      that align with precession. The dashed lines in \textbf{B} show all
      orbital frequencies aligning with frequencies in the ice volume data
      except for the 400.1\,kyr peak in eccentricity.}
\label{fig:orbital_and_benthic_time_series_power_specs}
\end{figure}

In this paper, we show that ice volume dynamics can mostly be explained
by a linear OFPA model without the use of a switching mechanism. Although our
model supports the approach, OFPA has two commonly cited limitations. Firstly,
eccentricity is the only parameter that oscillates with a period of around
100\,kyr. However, as we show in Sec. \ref{sec:orbital_params}, it can only
vary the magnitude of annual insolation by 0.2\%. This has led some to argue
that an Earth based amplification mechanism is necessary in order to explain the
100\,kyr period~\cite{imbrie_inertia,saltzman,nonlinear_amplifier}. Another
limitation of the OFPA approach relates to the second prominent frequency of
eccentricity. As shown in
Fig.~\ref{fig:orbital_and_benthic_time_series_power_specs}, the eccentricity
signal also contains a period of 400\,kyr. This is not clearly discernible in
the ice volume data, leading some studies to suggest that the 100\,kyr period
instead results from the interplay of the other obliquity and precession \cite{obliquity_subharmonics,intrinsic_no_eccentricity,no_ecc_just_bet_and_rho}.

In this paper, we aim to address both of the previous issues, and in doing so,
provide support for the OFPA school of thought. We first address the notion that
amplification of the eccentricity signal is necessary to explain its prominence
in the ice volume data. Using a simple model of bulk ocean temperature, we
estimate the range of warming rates that eccentricity can induce. Even with
conservative parameter estimates, we find that the difference in rates is
sufficient to explain the magnitude of ocean temperature dynamics, without the
need for amplification.

We also propose that the 400\,kyr period of eccentricity is absent from the ice
volume data because, instead of its magnitude, it is the change in eccentricity
that forces the ice volume dynamics. We find that the change in eccentricity
contains a far weaker 400\,kyr period and could arise from two terrestrial
mechanisms that respond to eccentricity with different lags. We propose these
could be the slow changing ocean temperature and the fast changing surface air
temperature. If these mechanisms have opposing effects on ice volume, they can
combine to produce a signal that resembles the change in eccentricity over time.

These findings are used to develop a phenomenological model of ice volume,
comprising three instantaneous orbital parameter terms and a lagged eccentricity
term. To evaluate the importance of the model parameters, we systematically
prune them and refit the model whilst evaluating the performance each time. The
key finding from this is that both the lagged and instantaneous eccentricity
terms must be included in order to produce 100\,kyr cycles with a consistent
amplitude. Excluding either of these terms from the full model reduces the
accuracy by approximately 40\%, as is shown in Fig.
\ref{fig:leave_one_out_with_diagram}.

We propose a physical interpretation for the components of our phenomenological
model, whereby ocean temperature produces the lagged eccentricity signal and the
3 instantaneous orbital parameters describe the surface air temperature. Using
our already fit parameter values, along with some extra constraints from data,
we are able to model the dynamics of ice volume, ocean temperature, and surface
air temperature. Each variable is compared with the equivalent proxy data over
the past 800\,kyr, showing reasonable agreement for most of the time
period, supporting the validity of these intermediate physical variables.
\section{Orbital Parameters}
\label{sec:orbital_params}
\begin{figure}
  \centering
  \includegraphics[width=0.8\linewidth]{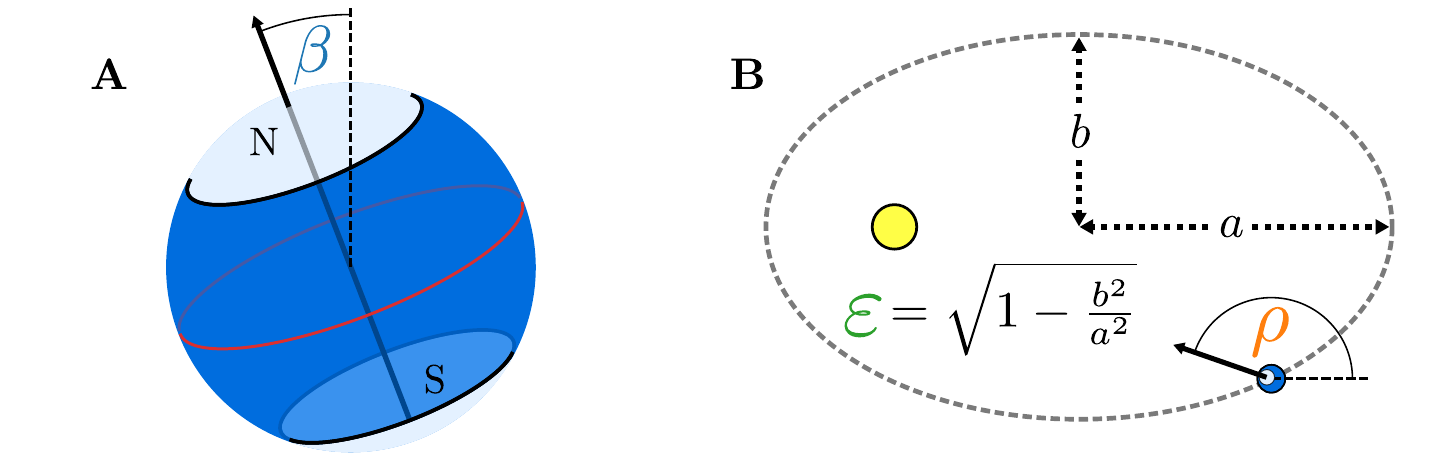}
  \caption{Diagrams of Earth defining the three orbital parameters. \textbf{A}:
    obliquity $\beta$ describes the tilt of Earth's rotational axis from
    vertical in the ecliptic frame. \textbf{B}: Eccentricity $\varepsilon$ is a
    function of the semi-major and semi-minor axes of Earth's orbit and has been
    exaggerated here for the purpose of visualisation. Precession $\rho$
    describes the angle of Earth's rotational axis around the vertical axis in
    the ecliptic frame, measured from the aphelion.}
  \label{fig:orbital_params}
\end{figure}

Our model uses Earth's three orbital parameters as inputs, the source for which
is discussed in Appendix \ref{sec:orbital_data}. These are obliquity
$\beta$, precession $\rho$, and eccentricity $\varepsilon$, which are
shown in Fig.~\ref{fig:orbital_params}. As obliquity increases, it
redistributes insolation away from the equator towards the poles. Precession
is the angle of Earth's rotational axis projected in the ecliptic plane and
causes a larger seasonal difference in insolation for one hemisphere, whilst
reducing the difference in the other. To better reflect the sinusoidal impact
that this angle has on insolation, we use the cosine of precession in our model.
Unlike obliquity and precession, eccentricity changes the total insolation
received by Earth over a year.

Glaciers spread down from the poles over land masses, and therefore exhibit more
variation in the land-dense northern hemisphere. Due to this asymmetry, it is
logical that the redistribution of insolation from obliquity and precession
would be able to impact ice volume variation, despite not changing the total
amount of insolation. Evidence of this impact is shown by the power spectra in
Fig. \ref{fig:orbital_and_benthic_time_series_power_specs}. As obliquity
increases, it exposes the polar ice to more insolation, so we expect ice volume
to be inversely related to obliquity. Since we are measuring precession
from the aphelion of Earth's orbit, the maximum of its cosine coincides with
maximum intensity insolation during the northern hemisphere summer. Milankovitch
proposed that $Q_{65}$ correlates with glacial retreat, which would mean ice
volume is inversely related to the cosine of precession~\cite{milankovitch}.

To understand how eccentricity affects insolation, we use the derivation
from McGehee and Lehman~\cite{insol_latlon}. This gives the annually averaged
insolation over Earth's surface as
\[
  \overline{Q_\mathrm{E}}(\varepsilon(t)) = \frac{K}{16\pi
  a^2\sqrt{1-\varepsilon(t)^2}},%\numberthis
  \label{eq:annual_insol}
\numberall
\]
where $\varepsilon(t)$ is eccentricity over time, shown in Fig.
\ref{fig:orbital_and_benthic_time_series_power_specs}, $a$ is the semi-major
axis of Earth's orbit, and $K$ is the solar constant, both of which are given in
Table \ref{tab:constants}. Note that because this is annually averaged, it is
independent of both obliquity and precession.

The Earth's orbital eccentricity varies between approximately 0 and 0.06, as
shown in Fig. \ref{fig:orbital_and_benthic_time_series_power_specs}. If we
substitute these limits into $\overline{Q_\mathrm{E}}$, along with the
constants given in Table \ref{tab:constants}, we find the annually averaged
insolation ranges from 340.353 to 340.967\,W/m$^2$. This corresponds to a
maximum change of 0.18\% due to eccentricity.

This seemingly insignificant variation has led some researchers to propose the
need for an Earth-based amplifier in order for eccentricity to have any
significant impact on ice volume
\cite{imbrie_inertia,saltzman,nonlinear_amplifier}. Although we do not rule out
the existence of these amplifiers, it is also true that the 100\,kyr period of
eccentricity allows for a prolonged increase or decrease in insolation to span
many thousands of years. This could potentially produce gradual changes that
build up in the Earth system, resulting in a significant impact on ice
volume. We propose that a mechanism that is slow to equilibrate, such as
ocean temperature, could partially store the excess energy produced as eccentricity
increases, allowing for significant warming to occur over thousands of years.

We propose that eccentricity is capable of explaining the broad dynamics of
ocean temperature, as observed in the paleoclimate record shown in Fig.
\ref{fig:sea_and_air_temp_with_map}. In order to test this hypothesis, we
estimate how eccentricity affects the rate of change of ocean temperature and
evaluate if there is enough energy variation in the system to produce the
observed range in ocean temperature.

In this derivation, we exclude the surface ocean since its interface with
the atmosphere causes it to behave differently than the rest of the ocean. We
will instead consider the bulk ocean temperature~$O(t)$, relating to all water
below a certain depth. We also treat the heat loss rate of the bulk ocean as
constant over time, however we will improve upon this assumption in Sec.
\ref{sec:fitting}.

The rate of change of the bulk ocean temperature is given by
\[
  \frac{\mathrm{d}O}{\mathrm{d}t} = \frac{\text{Total Absorbed Power}}{\text{Ocean
  Heat Capacity}} - \text{Heat Loss Rate}.
\numberall
\]
The total power provided to the bulk ocean through insolation can be expressed as
\[
  P_O(\varepsilon(t)) = \alpha \gamma A_O \overline{Q_\mathrm{E}}(\varepsilon(t)),
\numberall
\]
which has units W. The $\alpha$ coefficient accounts for the absorption and
reflection that occurs whilst passing through the atmosphere. The coefficient
$0<\gamma\leq1$ accounts for the heat emitted back to the atmosphere from the
surface ocean layer before it can reach the bulk volume of the ocean, where
$\gamma=1$ implies no loss of heat to the atmosphere. The annually averaged
insolation reaching Earth $\overline{Q_\mathrm{E}}$ comes from
\eqref{eq:annual_insol}, which is then scaled by the total area of the ocean
$A_O$. For this approximate calculation we are ignoring the latitudinal
asymmetry of ocean water, treating it as uniformly distributed across Earth's
surface. All of these values, apart from the free parameter $\gamma$, are given
in Table \ref{tab:constants}.

This power is then divided by the total heat capacity of the ocean, given as
\[
  C_O=\rho_OV_Oc_O,
\numberall
\]
which has units J/$^\circ$C. The mass of the ocean is given by $\rho_OV_O$,
where $\rho_O$ is the average density of the ocean and $V_O$ is the total
volume. Although heat capacity varies with the salinity and temperature of the
ocean, we will express the average specific heat capacity with $c_O$.

We can now substitute in our measured values to estimate the bulk ocean warming
rate. Since this will have units of $^\circ$C/s, we rescale by
$3.1536\times10^{10}$ to attain the rate in $^\circ$C/kyr.

This gives us a warming rate of
\[
  \frac{\mathrm{d}O}{\mathrm{d}t} = \frac{P_O(\varepsilon(t))}{C_O} - l=
  \frac{353.907\gamma}{\sqrt{1-\varepsilon(t)^2}} - l
  {}\,^\circ\mathrm{C/kyr},%\numberthis
  \label{eq:bulk_heating_rate}
\numberall
\]
where $l$ is the constant heat loss term and $\gamma$ is the absorption
constant. Since we have not formally defined the depth at which our bulk ocean
meets the ocean surface, we do not have a value for $\gamma$. Instead, we treat
this, and $l$ as free parameters to be determined through fitting. If
eccentricity is capable of producing the observed ocean temperature changes
without amplification, then we would expect a fit value of $\gamma<1$. A fit
value of $\gamma>1$ is not physically valid and would suggest that there is some
amplification mechanism that we have not accounted for.
\begingroup
\renewcommand{\arraystretch}{1.1}
\begin{table}
  \centering
\caption{Constants used in this paper.}
\label{tab:constants}
\begin{tabular}{c|c|c|c|c}
Term                    & Symbol     & Value                 & Units     & Source\\ \hline\hline
Solar Luminosity        & $K$        & $3.8287\times10^{26}$ & W         & \cite{solar_constant}   \\ \hline
Semi-Major Axis         & $a$        & $1.4960\times10^{11}$ & m         & \cite{solar_constant}  \\ \hline
Ocean Surface Absorption Ratio  & $\alpha$   & $0.48$                & 1         & \cite{ocean_heat_budget}  \\ \hline
Average Ocean Density   & $\rho_O$   & $1025$                & kg/m$^3$  & \cite{ocean_spec_heat}  \\ \hline
Ocean Specific Heat     & $c_O$      & $3850$                & J/kg\,$^\circ$C  & \cite{ocean_spec_heat}  \\ \hline
Ocean Volume            & $V_O$      & $1.335\times10^{18}$  & m$^3$     & \cite{ocean_vol}  \\ \hline
Ocean Surface Area      & $A_O$      & $3.619\times10^{14}$  & m$^2$     & \cite{ocean_vol}  \\ \hline
Ocean Heat Capacity     & $C_O$      & $5.268\times10^{24}$  & J/$^\circ$C&\cite{ocean_spec_heat,ocean_vol}  \\ \hline
Average Eccentricity    & $\overline{\varepsilon}$ & 0.02707 & 1         & \cite{laskar2004} \\ \hline
Average Obliquity       & $\overline{\beta}$       & 0.40739 & rad       & \cite{laskar2004} \\ \hline
Average Ocean Input Power& $P_O(\overline{\varepsilon})$     & $5.914\times10^{16}$ & J/s
&\cite{laskar2004,ocean_heat_budget,solar_constant,ocean_vol}
\end{tabular}
\end{table}
\endgroup

As a simple calculation, we explore the role $\gamma$ plays in this warming rate,
using ocean temperature data as a reference. The last glacial maximum occurred
around 20\,kya, and since then our ocean has heated at an extreme rate, similar 
to other transitions into an interglacial period~\cite{lgm}. The surface water
temperature (SWT) has increased by up to 10$^\circ$C in individual
locations~\cite{sst_temp_ranges}. However, globally averaged SWT and bottom
water temperature (BWT) have increased by around 3 to 5$^\circ$C
respectively~\cite{benthic_to_ice_vol, sst_lgm_map}. This gives us an
approximate warming rate of 0.2$^\circ$C/kyr.

If we suppose that half of the incoming heat reaches the bulk ocean, meaning
that $\gamma=0.5$, then for an eccentricity range of $0\leq\varepsilon\leq0.06$,
\eqref{eq:bulk_heating_rate} would give a difference in warming rates of
0.319$^\circ$C/kyr. In Sec. \ref{sec:fitting}, we find from fitting that
$\gamma=0.59$, resulting in a similar rate as calculated here, and also of the
same scale as the 0.2$^\circ$C/kyr we see in the data. This suggests that
eccentricity could explain the observed ocean temperature changes without the
need for amplification.

It is important to note that we have used the extrema of eccentricity and ocean
temperature for this calculation. We also acknowledge that the ocean heat loss
rate will vary as a function of temperature to some degree. However, this simple
calculation shows that, given the range for eccentricity, it has the ability to
drive a change in ocean temperature on the same scale as that observed in the
data.

\section{Phenomenological Model}
\label{sec:phenom_model}
We now present a simple phenomenological model that is intended to reproduce the
global ice volume data using only a linear function of the orbital parameters.
We then fit the coefficients of this model to the data and evaluate its
predictive power. Although we do not expect this model to outperform the more
complex, non-linear ice volume models, it will demonstrate the extent to which a
linear OFPA model can approximate the ice volume dynamics. We justify the
inclusion of each term in our model by refitting it with every subset of terms,
evaluating the degree to which it can explain the data in each case.

\begin{figure}
  \centering
  \input{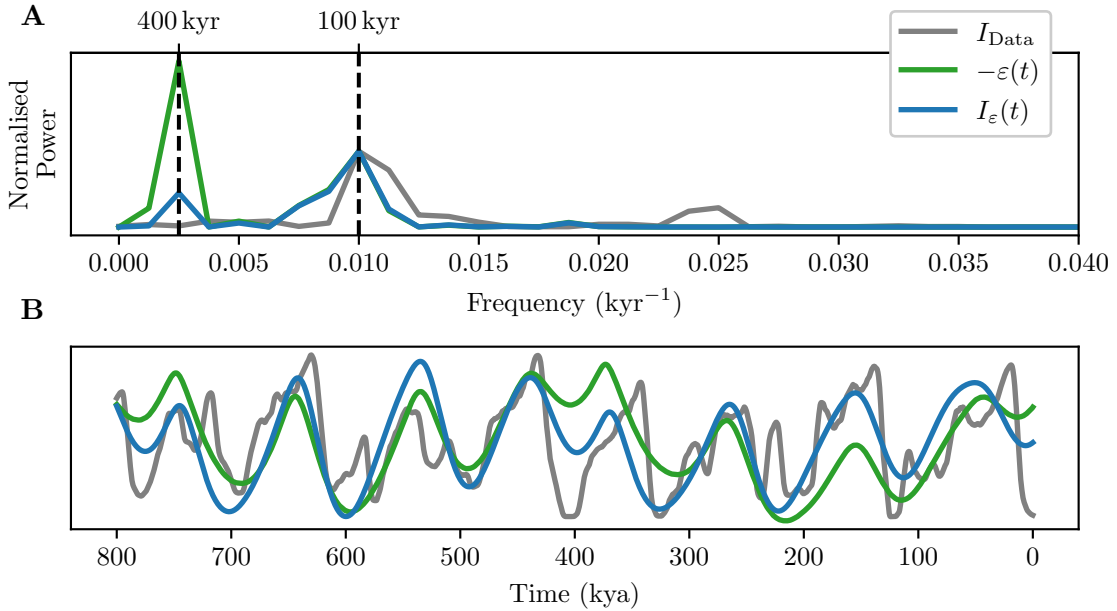}
  \caption{\textbf{A}: Power spectra comparison of the ice volume data,
    the negative eccentricity curve $-\varepsilon(t)$, and the eccentricity
    component of our model solution $I_\varepsilon(t)$. They have been
    normalised to equate the powers corresponding to 100\,kyr. Our
    $I_\varepsilon(t)$ power spectrum matches that of $\varepsilon(t)$
    apart from a significant drop around the 400\,kyr period. \textbf{B}: Time
    series showing the same comparison, also normalised for qualitative
    comparison. They show how the filtered eccentricity better captures the
    broad behaviour of the ice volume curve, namely around 780, 400, 320, and
    160\,kya.}
  \label{fig:deps_approx_power_spec}
\end{figure}

The model is based on the assumption that the change in ice volume depends
instantaneously on the three orbital parameters, as well as a lagged version of
eccentricity. The instantaneous terms allow us to reproduce the orbitally
aligned power spikes in the ice volume data from Fig.
\ref{fig:orbital_and_benthic_time_series_power_specs}. The lagged term interacts
with the instantaneous eccentricity term of the opposite sign to approximate the
change in eccentricity. This effectively removes the 400\,kyr period from the
ice volume solution to better align with the data, as shown in Fig.
\ref{fig:deps_approx_power_spec}.

To achieve this, the model comprises two linear differential equations,
describing the evolution of ice volume $I(t)$, and the slow-responding variable
$\tilde{\varepsilon}(t)$, which resembles the eccentricity signal with a lag of
approximately $\tau$. This is because $\tilde{\varepsilon}$ changes according to
the difference between itself and eccentricity $\varepsilon$, converging towards
$\varepsilon$ at a rate determined by $\tau$. If $\tau=0$ then
\eqref{eq:phenom_model_slow_var} would simplify to
$\tilde{\varepsilon}(t)=\varepsilon(t)$, meaning there is no lag between the two
signals. However, as $\tau$ increases, $\tilde{\varepsilon}$ converges more
slowly towards $\varepsilon$, and so the lag between the two signals increases.
Although we have suggested that ocean temperature could be this slow-responding
variable, here we are purely concerned with reproducing the ice
volume data with no physical interpretation or additional assumptions.

The model is therefore given by
\begin{align}
  \tau\frac{\mathrm{d}\tilde{\varepsilon}}{\mathrm{d}t}
  &= \varepsilon(t) - \tilde{\varepsilon}(t),
  \label{eq:phenom_model_slow_var}\\[4pt]
  \tau\frac{\mathrm{d}I}{\mathrm{d}t} &= p_1 \tilde{\varepsilon}(t) -  p_2\varepsilon(t) -
  p_3\beta(t) - p_4\cos(\rho(t)) - I(t) + p_5,%\numberthis
  \label{eq:phenom_model_ode}
\end{align}
where the $p_i$ coefficients are to be fitted, along with the time
constant $\tau$. This is introduced since we do not expect ice volume, nor the
slow variable, to respond instantaneously to changes in the orbital parameters.
Two separate time constants were originally used for this model, however,
fitting revealed the two constants to be approximately equal. This could be due
to some coupling between ice volume and the mechanism that $\tilde{\varepsilon}(t)$
represents, causing them to change at the same rate.

We now present the analytical solution to this model and fit the coefficients to
the ice volume data. Using the integrating factor method, we solve for the
slow variable to get
\[
  \tilde{\varepsilon}(t) = \zeta_\tau[\varepsilon(t)] +
  \tilde{\varepsilon}_0e^{\nicefrac{-t}{\tau}},
\numberall
\]
where $\tilde{\varepsilon}_0$ is the initial condition and the functional
\[
  \zeta_\tau[y(t)] = \frac{e^{\nicefrac{-t}{\tau}}}{\tau}\int_0^t y(u)
  e^{\nicefrac{u}{\tau}}\mathrm{d}u,%\numberthis
  \label{eq:zeta_functional}
\numberall
\]
for some function $y(t)$.

Substituting this into the differential equation for ice volume gives
\[
  \tau\frac{\mathrm{d}I}{\mathrm{d}t} = p_1\zeta_\tau[\varepsilon(t)] -
  p_2\varepsilon(t) - p_3\beta(t) - p_4\cos(\rho(t)) + p_5 +
  p_1 \tilde{\varepsilon}_0e^{\nicefrac{-t}{\tau}}.%\numberthis
\numberall
\label{eq:phenom_model_ode_subbed}
\]
We then solve this using the integrating factor method again to get
\[
  I(t) = p_1\zeta_\tau\big[\zeta_\tau[\varepsilon(t)]\big] -
  p_2\zeta_\tau[\varepsilon(t)] - p_3\zeta_\tau[\beta(t)] -
  p_4\zeta_\tau[\cos(\rho(t))] + p_5 + \left(\frac{p_1 \tilde{\varepsilon}_0 t}{\tau}
  + I_0 - p_5\right)e^{\nicefrac{-t}{\tau}}.
\numberall
\]

\begin{figure}
  \centering
  \input{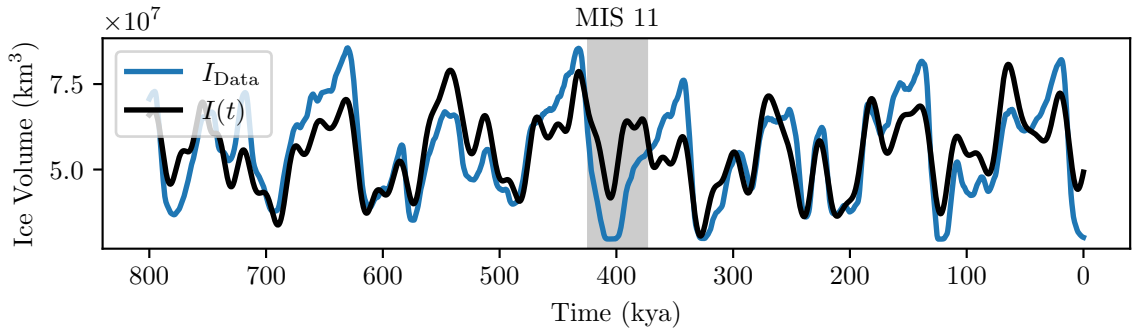}
  \caption{Our modelled ice volume $I(t)$ from \eqref{eq:phenom_model_sol},
    alongside the ice volume data $I_\mathrm{Data}$. The grey region delineates
    Marine Isotope Stage (MIS) 11, around which there is a notable difference
    between the two curves. The model parameters that produce this fit are given
    in Table \ref{tab:phenom_params}.}
  \label{fig:fit_ice_vol}
\end{figure}

If we run our model for sufficiently long before our period of interest, the
$e^{\nicefrac{-t}{\tau}}$ term in $I(t)$ can be treated as zero. This gives the
asymptotic approximation as
\[
  I(t) = p_1\zeta_\tau\big[\zeta_\tau[\varepsilon(t)]\big] -
  p_2\zeta_\tau[\varepsilon(t)] - p_3\zeta_\tau[\beta(t)] -
  p_4\zeta_\tau[\cos(\rho(t))] + p_5.%\numberthis
  \label{eq:phenom_model_sol}
\numberall
\]
This solution is non-linear in $\tau$, so to optimise the parameters of the
model, we repeat a least squares fit for the $p_i$ coefficients whilst varying
$\tau$, guaranteeing that the parameters are globally optimised. The optimal
parameter values for this model are given in Table \ref{tab:phenom_params}, with
the corresponding solution shown in Fig. \ref{fig:fit_ice_vol}.
\begin{table}
  \centering
  \caption{Parameter values for the phenomenological model given in
    \eqref{eq:phenom_model_ode}. Their roles are shown in
  Fig.~\ref{fig:leave_one_out_with_diagram}B. Errors given by the 95\%
confidence interval in the fit are shown beside the estimated value.}
\label{tab:phenom_params}
\begin{tabular}{c|c|c|c}
  Parameter & Role & Value & Units               \\ \hline\hline
  $p_1$  & Slow Eccentricity &$1.88\pm0.05$&$\times10^9$\,km$^3$ \\ \hline
  $p_2$  & Fast Eccentricity &$1.94\pm0.05$&$\times10^9$\,km$^3$\\ \hline
  $p_3$  & Fast Obliquity &$1.54\pm0.06$&$\times10^9$\,km$^3$\\ \hline
  $p_4$  & Fast Precession  &$1.9\pm0.1  $&$\times10^7$\,km$^3$\\ \hline
  $p_5$  & Constant Offset &$6.8\pm0.2  $&$\times10^8$\,km$^3$ \\ \hline
  $\tau$ & Response Rate &$14.8\pm0.4 $& kyr
\end{tabular}
\end{table}

The ice volume solution in Fig. \ref{fig:fit_ice_vol} shows reasonably close
agreement with the data, with the notable exception of the time interval around
Marine Isotope Stage (MIS) 11, which will be discussed in Sec.
\ref{sec:discussion}. We emphasise that this model is not intended to be a
perfect fit to the data, but rather a demonstration of how much can be explained
by a linear dependence on the orbital parameters. By reproducing the majority of
the ice volume data with this linear model, we propose that more complex
mechanisms are only needed to explain the few remaining extrema such as MIS 11.

\begin{figure}
  \centering
  \input{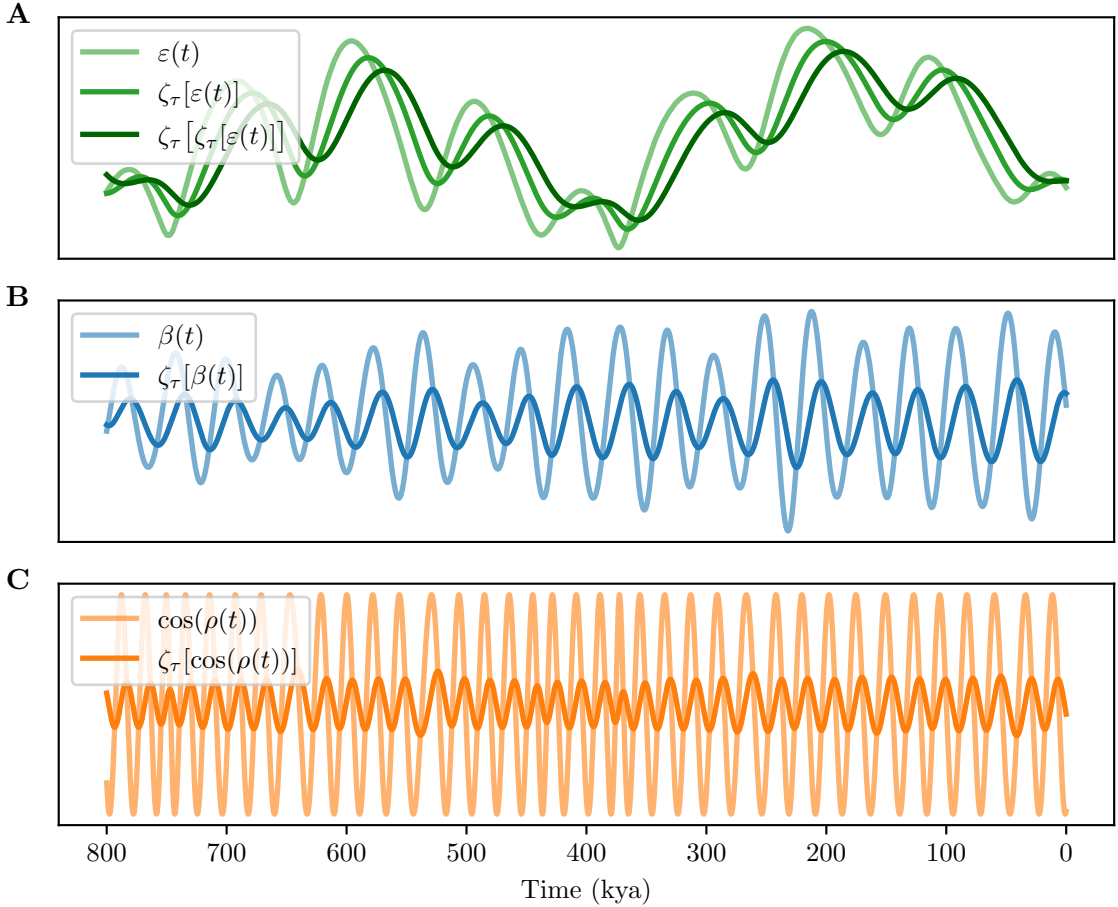}
  \caption{The qualitative effect of the functional $\zeta_\tau$ on eccentricity
    (\textbf{A}), obliquity (\textbf{B}), and the cosine of precession
    (\textbf{C}) as defined by \eqref{eq:zeta_functional}, where
    $\tau=14.8$\,kyr. Each $\zeta_\tau$ term that appears in
    \eqref{eq:phenom_model_sol} is shown. Note how the higher the
    frequency of the orbital parameter, the more $\zeta_\tau$ reduces its
    amplitude.}
  \label{fig:zeta_impact}
\end{figure}

The analytical solution for ice volume is expressed in terms of the $\zeta_\tau$
functional. To demonstrate how $\zeta_\tau$ responds to its inputs,
Fig.~\ref{fig:zeta_impact} shows each term in \eqref{eq:phenom_model_sol}
plotted alongside the orbital parameter it depends on. We see that the effect of
$\zeta_\tau$ is somewhat different in each case. For eccentricity, a lag
approximately equal to $\tau=14.8$\,kyr is introduced each time $\zeta_\tau$ is
applied. For obliquity and precession, the amplitude is reduced more
significantly, and the lag is shorter. This is because they oscillate at a
higher frequency than eccentricity. As a result, where the
$\zeta_\tau[\varepsilon(t)]$ curve is able to slowly follow the $\varepsilon(t)$
curve, $\zeta_\tau[\beta(t)]$ and $\zeta_\tau[\cos(\rho(t))]$ cannot reach the
extrema of their inputs before they begin to change direction. However, since
the scale factors in \eqref{eq:phenom_model_sol} will account for this effect,
the only important feature of $\zeta_\tau$ is the lag it introduces. The ice
volume solution can therefore be interpreted simply as a weighted sum of the
lagged orbital parameters, where eccentricity appears twice, with a longer lag
the second time.

\begin{figure}
  \hspace{-30pt}
  \input{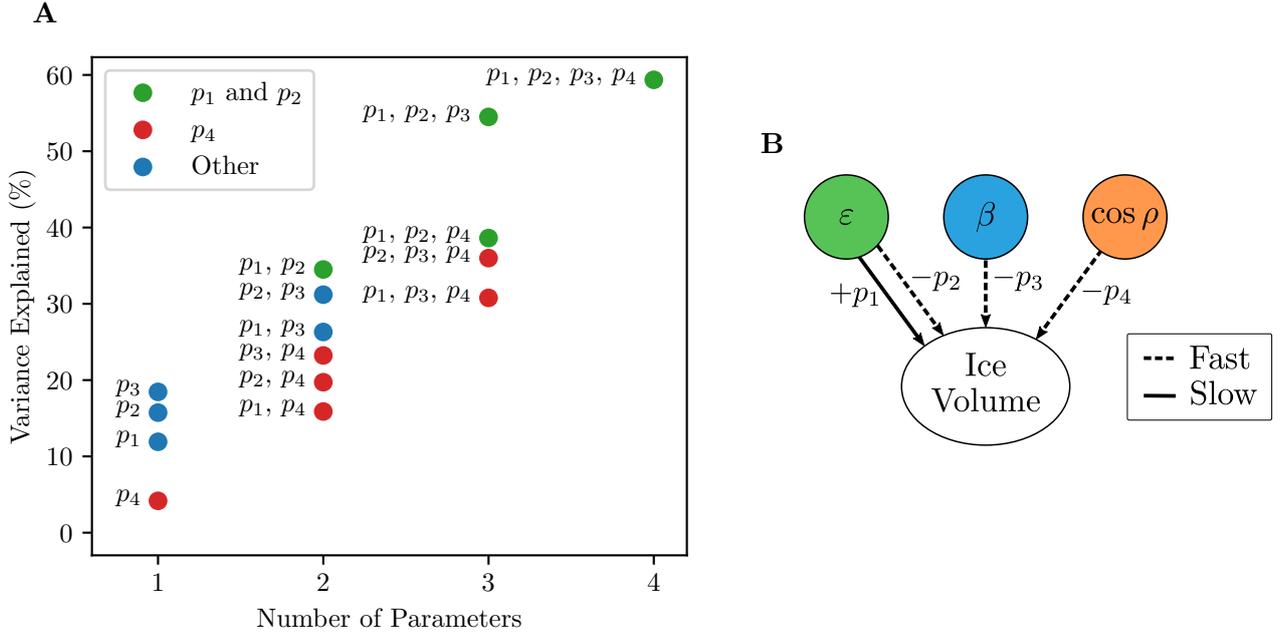}
  \caption{\textbf{A}: Variance explained for all possible parameter
    combinations of the model given by \eqref{eq:phenom_model_ode}, with the
    excluded parameters set to zero. The constant term $p_5$ and time constant
    $\tau$ are always included. For each case, the included parameters were
    optimised to attain the best fit to the ice volume data. Cases where both
    $p_1$ and $p_2$ are included (green) produce especially good fits, whilst
    cases without this pair, but with $p_4$ included (red), produce especially
    poor fits. \textbf{B}: Flow diagram showing the role of each parameter in
    the model. The fast arrows (dashed) represent an instantaneous response,
    scaled by the accompanying parameter, whilst the slow arrows (solid) depict
    a response to the input that is scaled by the accompanying parameter and
    lagged by approximately $\tau$.}
    \label{fig:leave_one_out_with_diagram}
\end{figure}

In order to evaluate the necessity of each term in our model, we systematically
prune terms whilst evaluating the accuracy in each case. The results of this are
shown in Fig.~\ref{fig:leave_one_out_with_diagram}. To measure the accuracy,
we are using the variance of the ice volume data that is explained by the model.
The offset term $p_5$ and time constant $\tau$ are included in every version of
the model so that the solutions are comparable. The precession term, represented
by $p_4$, is shown to consistently contribute less to the variance explained
than the other terms. This is partly due to the nature of the ice volume curve,
in which the higher frequencies appear with smaller amplitudes. However, this
may also arise due to solid Earth forcing, such as volcanic activity, that could
affect the ice volume on this timescale, introducing noise that precession
cannot account for. Although precession contributes the least, it is still
responsible for approximately 5\% of the variance in the ice volume data,
regardless of the other terms included in the model. This suggests that it is
not contributing to over-fitting. Moreover, its inclusion allows for the higher
frequencies, shown in Fig. \ref{fig:benth_and_power_specs}, to be represented
in the model.

Secondly, we find that the subsets that include both $p_1$ and $p_2$,
relating to the slow and fast eccentricity terms, consistently produce the best
fit. This indicates that the change in eccentricity represented by the pair is
crucial to the model, more so than either term on its own.
In order to better understand the role of the slow variable
$\tilde{\varepsilon}(t)$, we examine how eccentricity relates to ice volume. As
shown in Fig.~\ref{fig:orbital_and_benthic_time_series_power_specs}, ice
volume decreases during peaks of eccentricity, which is consistent with
\eqref{eq:annual_insol}. However, the degree to which ice volume decreases
appears to be independent of the magnitude of eccentricity's peaks. This
suggests that the absolute value of eccentricity may be less important than the
direction in which it is changing. By including the $\tilde{\varepsilon}(t)$
variable alongside $\varepsilon(t)$, we are providing the model with both
current, and lagged, values of eccentricity. The difference of these two signals
represents the change in eccentricity over time, shown by the first two terms in
\eqref{eq:phenom_model_ode_subbed}.

As shown in Table \ref{tab:phenom_params}, the fit $p_1$ and $p_2$ values
are within 3\% of each other. This indicates that the optimised model only depends
on the change in eccentricity and not the instantaneous or lagged values on
their own. We could therefore simplify the model by setting $p_2=p_1$, without
losing accuracy. However, since we wish to find physical interpretations for
both the slow and fast responses to eccentricity we will keep the two terms
distinct.

By using the relatively short-term change in eccentricity as input, rather than
eccentricity itself, our model is able to produce 100\,kyr oscillations whilst
effectively removing the 400\,kyr amplitude modulation.
Fig.~\ref{fig:deps_approx_power_spec}A shows the prominent 400\,kyr peak in
eccentricity's power spectrum that is significantly reduced in the
eccentricity component of our model solution $I_\varepsilon(t)$.
Fig.~\ref{fig:deps_approx_power_spec}B shows this effect in the time domain,
with $I_\varepsilon(t)$ more closely matching the broad dynamics of the ice
volume data.

\section{Physical Interpretation}
\label{sec:physical_interpretation}
So far, we have presented a phenomenological model that is able to capture the
broad dynamics of the ice volume data. We now propose a physical interpretation
for the terms of this model and attempt to produce solutions for these physical
mechanisms that are consistent with the relevant data.

\subsection{Choice of Mechanisms}
\label{sec:choice_of_mechanisms}
The phenomenological model contains three terms that respond quickly to the
orbital parameters, and one that responds slowly. We propose that these slow and
fast dynamics are attributed to separate mechanisms. The fast part of the model
could represent the response of Earth's surface air temperature (SAT) to the
orbitally varied insolation, which can be considered instantaneous on this
timescale. As discussed in Sec. \ref{sec:orbital_params}, an increase in
eccentricity, obliquity, or precession can reduce ice volume. This predominantly
occurs through SAT and direct radiation of the glaciers. As these are closely
correlated effects, and proxy data only exists for SAT, we treat this as a
combined surface temperature variable $S(t)$ and use SAT data as a proxy for its
qualitative behaviour.

We propose that the slow variable $\tilde{\varepsilon}(t)$ relates to ocean
temperature $O(t)$ in some way. Due to its high thermal inertia, it could take
thousands of years for the ocean to equilibrate to a new temperature. This
response time is slow enough that the impact of precession and obliquity are
significantly reduced, as can be seen in Fig.~\ref{fig:zeta_impact}. To
confirm this, we trialled an ocean model with all three orbital parameters
and found no significant improvement to the accuracy of the fit.

Although we are treating $O(t)$ as ocean temperature, it may also incorporate
mechanisms such as long-term feedbacks in atmospheric CO$_2$. The ocean is one
of the largest carbon sinks on Earth, but holds less CO$_2$ at higher
temperatures~\cite{hot_ocean_loses_co2}. This positive correlation between ocean
temperature and atmospheric CO$_2$ concentration means it is difficult to
separate the impact of each on ice volume. However, as with SAT and radiation,
these effects are closely correlated, and so we choose to represent these
correlated mechanisms with our single slow variable $O(t)$. This allows us to
use ocean temperature data as a proxy for its qualitative behaviour.

We are using the fit parameters from our phenomenological model, which means we
are assuming ocean temperature has a time constant of $\tau=14.8$\,kyr.
However, since the temperature of the ocean never reaches a global equilibrium,
it is difficult to determine if this is physically plausible. Van Aken states
that heat can be mixed across all depths on a time scale of hundreds to a few
thousand years~\cite{ocean_100_to_1000_timescale}. The Intergovernmental Panel
on Climate Change state that deep ocean temperatures can take thousands of years
to respond to surface temperature changes~\cite{ocean_1000_timescale}. Crucifix
estimates an even longer mixing time on the order of
10\,kyr~\cite{crucifix_original}. There is a similar level of uncertainty when
it comes to estimating the time constant for ice sheet growth and ablation.
Estimates range from 500 years on the lower end~\cite{ice_time_const_short}, but
can go up to 27\,kyr~\cite{ice_time_const_long}. Our fit time constant of
$\tau=14.8\pm0.4$\,kyr therefore seems feasible for both ocean temperature and
ice volume.

Although we have demonstrated that the observed warming rate of the ocean can be
sufficiently explained by eccentricity, we have so far treated the heat
loss rate as constant. Heat loss occurs through back radiation, convection, conduction, and
evaporation~\cite{ocean_heat_budget}. We will assume for simplicity that these
are all linearly dependent on bulk ocean temperature $O(t)$, giving a
consolidated loss term of $m O(t) + n$. Replacing the constant loss rate in
\eqref{eq:bulk_heating_rate} with this linear loss rate gives
\[
  \frac{\mathrm{d}O}{\mathrm{d}t} = \frac{353.907\gamma}{\sqrt{1-\varepsilon(t)^2}} - mO(t) - n,
\numberall
\]
which has units $^\circ$C/kyr.

As we wish to substitute $O(t)$ into our linear model, we take the first order
Taylor expansion about the average eccentricity $\overline{\varepsilon}$ to get
\[
  \frac{\mathrm{d}O}{\mathrm{d}t} = 9.591\gamma\varepsilon(t) + 353.8\gamma -
  mO(t) - n, %\numberthis
  \label{eq:linear_approx_ocean_sol}
\numberall
\]
where $\overline{\varepsilon}$ is given in Table \ref{tab:constants}.

In order to simplify this equation for use in our physical model, we write it as
\[
  \tau \frac{\mathrm{d}O}{\mathrm{d}t} = c\varepsilon(t) - O(t) + \alpha_O,
\numberall
\]
\vspace{-5pt}
where
\begin{align}
    \tau &= \frac{1}{m},\\[2pt]
    c &= \frac{9.591\gamma}{m},\\[2pt]
    \alpha_O &= \frac{353.8\gamma - n}{m}.
\end{align}
Once we have estimated values for $c$, and $\alpha_O$, we can estimate all of
the coefficients in \eqref{eq:linear_approx_ocean_sol}. For this equation to be
physically plausible, we would expect the resultant heat gain and heat loss
rates to be approximately equal, so as to avoid unbounded temperature change. We
also require that $0<\gamma\leq 1$ is satisfied.

\subsection{Physical Model}
Our physical model will produce the same ice volume $I(t)$ solution as the
phenomenological model, but will also produce solutions representing bulk ocean
temperature $O(t)$ and SAT $S(t)$. These three variables can then be compared to
their respective proxy data to help validate the model. To remain consistent
with the phenomenological model, we require that ocean temperature positively
impacts ice volume, whilst SAT negatively impacts it. It is logical that an
increase in SAT would lead to a decrease in ice volume, however it is less
obvious how an increase in ocean temperature could lead to an increase in ice
volume.

A possible explanation for this relates to the evaporation rate from the ocean
surface. As the ocean temperature increases, the evaporation rate also
increases, leading to an increase in the moisture content of the air. This could
then lead to greater precipitation over the ice sheets, increasing their volume.
It is difficult to determine the degree to which this would impact ice volume,
especially since ocean warming could also be attributed to a decrease in ice
volume. However, we emphasise that $O(t)$ does not necessarily represent global
ocean temperature, with one alternative being the ocean temperature only at high
latitudes. In this case, it is possible that increased moisture in the
atmosphere plays a more significant role.

\begin{figure}
  \centering
  \includegraphics[width=0.5\textwidth]{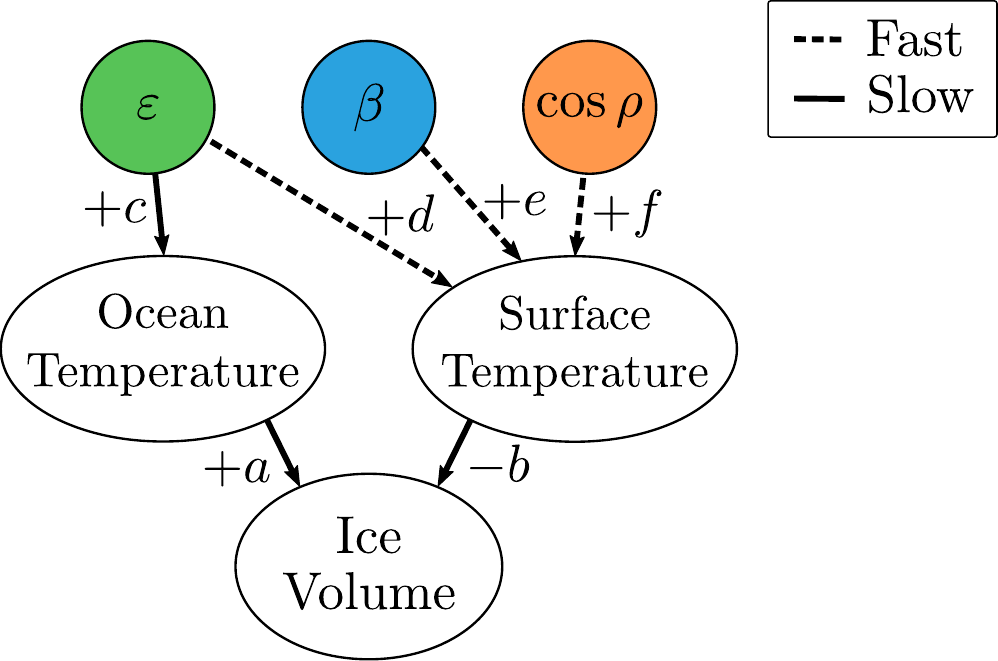}
  \caption{Flow diagram showing the proposed physical model of orbital influence
    through bulk ocean and surface air temperature. The fast arrows represent an instantaneous
    response, scaled by the accompanying parameter, whilst the slow arrows depict a
    response to the input that is scaled by the accompanying parameter and lagged by
    approximately $\tau$.}
  \label{fig:ode_diagram}
\end{figure}
The flow diagram in Fig.~\ref{fig:ode_diagram} shows this proposed physical
model, extending the phenomenological model from Sec. \ref{sec:phenom_model}.
The governing equations for this model are
\begin{align}
  \tau\frac{\mathrm{d} I}{\mathrm{d} t} &= a O(t) - b S(t) - I(t) +
  \alpha_I,%\numberthis
  \label{eq:ice_ode}\\
  \tau\frac{\mathrm{d} O}{\mathrm{d} t} &= c\varepsilon(t) - O(t) +
  \alpha_O,\\
  S(t) &= d\varepsilon(t) + e\beta(t) + f\cos\left({\rho(t)}\right) +
  \alpha_S,%\numberthis
  \label{eq:surface_steady}
\end{align}
where $a$, $b$, $c$, $d$, $e$, $f$, $\alpha_O$, $\alpha_I$, $\alpha_S$, and
$\tau$ are parameters that will be determined. In order to produce three
distinct solutions that can be compared to their relevant data, each variable
requires an offset and scaling factor. As a result, this physical model contains
10 parameters, compared to the 6 in the phenomenological model. In Sec.
\ref{sec:fitting}, we use the proxy data for ocean temperature and SAT to
introduce 4 extra constraints, allowing us to uniquely determine the physical
parameter values.

In order to analytically solve this system, we once again solve for the slow
variable $O(t)$ using an integrating factor. This is then substituted into the
ice volume equation, alongside the instantaneous $S(t)$. We then solve for
$I(t)$ using an integrating factor. The solution for $O(t)$ is given as
\[
  O(t) = c\zeta_\tau[\varepsilon(t)] + \alpha_O - \alpha_O
  e^{\nicefrac{-t}{\tau}}  + O_0e^{\nicefrac{-t}{\tau}},%\numberthis
  \label{eq:O_sol}
\numberall
\]
where $O_0$ is the initial condition for $O(t)$ and the functional
$\zeta_\tau$ is defined in \eqref{eq:zeta_functional}.

Substituting \eqref{eq:surface_steady} and \eqref{eq:O_sol} into
\eqref{eq:ice_ode} then gives
\[
  \tau\frac{\mathrm{d} I}{\mathrm{d} t} = a c\zeta_\tau[\varepsilon(t)]
  - b d \varepsilon{(t)} - b e \beta{(t)} - b f \cos{(\rho{(t)})}- I{(t)} + (a
  \alpha_{O} -  b\alpha_{S} + \alpha_{I}) + a(O_0 - \alpha_O)e^{\nicefrac{-t}{\tau}}.
\numberall
\]

The analytical solution for $I(t)$ is then
\begin{align}
  I(t) = ac\zeta_\tau\big[\zeta_\tau[\varepsilon(t)]\big] &- bd\zeta_\tau[\varepsilon(t)] -
be\zeta_\tau[\beta(t)] - bf\zeta_\tau[\cos(\rho(t))]  \\ &+(a \alpha_{O} -  b\alpha_{S} +
\alpha_{I}) + \left(\frac{a}{\tau}\big((O_0 - \alpha_O)t - \alpha_O \tau\big) +
b\alpha_S - \alpha_I + I_0\right)e^{\nicefrac{-t}{\tau}}.
\end{align}
As before, if we solve from sufficiently long before our period of interest, we
can asymptotically approximate this as
\begin{align}
  I(t) &= ac\zeta_\tau\big[\zeta_\tau[\varepsilon(t)]\big] - bd\zeta_\tau[\varepsilon(t)] -
be\zeta_\tau[\beta(t)] - bf\zeta_\tau[\cos(\rho(t))] + (a \alpha_{O} -  b\alpha_{S} +
\alpha_{I}),\\
     &= p_1\zeta_\tau\big[\zeta_\tau[\varepsilon(t)]\big] - p_2\zeta_\tau[\varepsilon(t)] -
     p_3\zeta_\tau[\beta(t)] - p_4\zeta_\tau[\cos(\rho(t))] + p_5,
\end{align}
where we have included the parameters from the phenomenological model from
\eqref{eq:phenom_model_sol} to demonstrate how its coefficients align with those
of the physical model. As is shown by this comparison, although the physical
model has 10 parameters, the two solutions for ice volume are still equivalent.
Where we originally had single $p_i$ coefficients scaling the orbital
parameters, we now have a product of two coefficients, one scaling the
intermediate physical variable and the other scaling its contribution to ice
volume. Similarly, the $p_5$ coefficient is now represented by the weighted sum
of three offsets, one for each physical variable. In order to understand the
physical quantities that we are modelling, we review the available proxy data
for each one and compare it to our model solutions.

\subsection{Ocean Temperature}
\begin{figure}
  \vspace{-42pt}
  \input{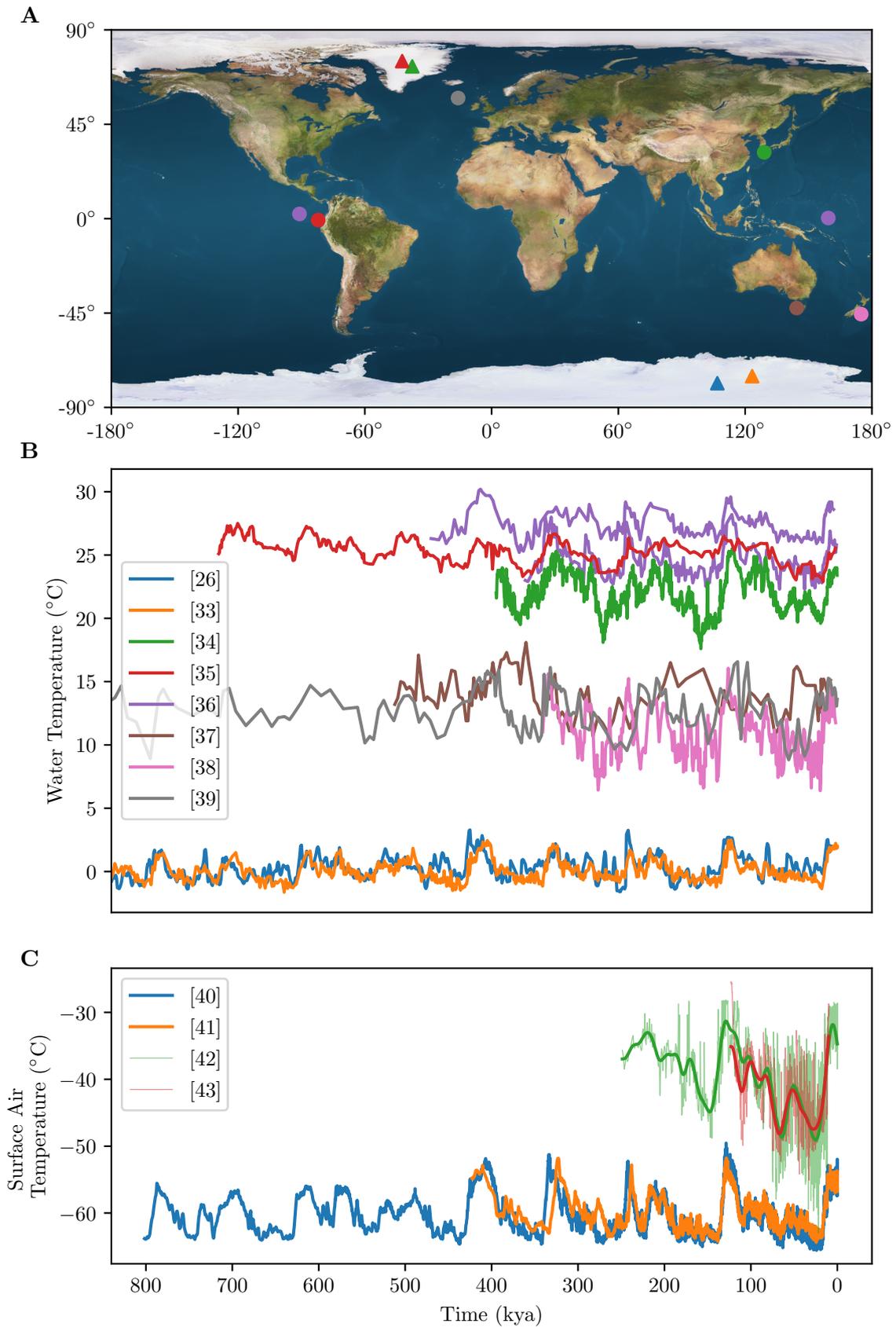}
  \caption{\textbf{A}: Drilled core locations used in our data comparison with
    seafloor drill sites shown as circles and ice core drill sites shown as
    triangles~\cite{earth_map}. \textbf{B}: Proxy data for BWT and SWT from
    independent drilled cores, their colours correspond to the locations shown
    on the map. The global BWT data are plotted in blue and orange with the
    remaining plots relating to the regionally dependent SWT data. \textbf{C}:
    Regional SAT for locations in Antarctica and Greenland. Since the Greenland
    data are more noisy, a second order low-pass Butterworth filter was used to
    remove frequencies above 19\,kyr$^{-1}$, as this is the highest frequency at
    which orbital parameters vary.}
  \label{fig:sea_and_air_temp_with_map}
\end{figure}

The ocean temperature time series shown in
Fig.~\ref{fig:sea_and_air_temp_with_map} shows both BWT and SWT proxy data.
The blue and orange plots represent the global BWT and show reasonable agreement
with each other. The blue plot comes from Elderfield, who used Mg/Ca ratios to
separate the BWT and ice volume contributions to benthic foraminifera
$\delta^{18}$O data~\cite{benthic_to_ice_vol}. The orange plot comes from
Shackleton's quadratic model of BWT from $\delta^{18}$O, fit using current
samples of foraminifera for which the local BWT is known~\cite{shack_bwt_eq}.
This model was then applied to the Lisiecki and Raymo $\delta^{18}$O data, which
largely comprises the same foraminifera used to calibrate Shackleton's
model~\cite{benthic_data}.

The remaining 7 plots relating to water temperature correspond to SWT proxy
data, which vary far more with location. Data sources were chosen to incorporate
drill sites from a range of latitudes, as shown on the accompanying map. The SWT
data are produced through a method similar to that used by Elderfield, with the
addition of a second measure that relies on alkenones. These are a type of
ketone that some phytoplankton produce, with more saturated alkenones being
produced in higher temperatures. Since phytoplankton live near the surface of
the water to collect sunlight, the degree of alkenone saturation found in these
phytoplankton can act as a proxy for SWT~\cite{alkenones}.

\subsection{Surface Air Temperature}
The third variable in our physical model represents globally averaged SAT. Proxy
data on the timescale we are interested in are limited, with the predominant
source being continental ice cores (Fig. \ref{fig:sea_and_air_temp_with_map}A).
As snow accumulates, the top layers slowly compress lower layers into ice,
preserving the isotopic composition of the snow at the time. The ice cores
containing the longest history are located in Antarctica and Greenland, reaching
depths of approximately 3\,km. Fig.~\ref{fig:sea_and_air_temp_with_map}C depicts
proxy SAT data for Greenland and Antarctica.

The blue and orange plots represent SAT at Antarctica's Dome C and Vostok
respectively. Both datasets were given as temperature anomalies from current
conditions, so we have shifted the plots to match with the present mean
temperature of -55$^\circ$C for this region of Antarctica~\cite{petit1999}. The
orange plot employs the same principle as is used to estimate global ice volume
from foraminifera shells. Since precipitation containing the lighter $^{16}$O is
more likely to evaporate, we expect to see greater proportions of $^{18}$O in
the preserved ice when temperatures were higher. Through experiment, Dansgaard
found this relationship to be approximately $T=1.45\delta + 19.7$, where
$\delta$ is the $\delta^{18}$O ratio~\cite{original_d18O_to_temp}. The blue plot
also uses this principle, but instead measures the ratio of hydrogen isotopes in
the ice. Although the isotopic ratios of hydrogen and oxygen are closely
related, they can deviate slightly due to the conditions in the region of ocean
from which the precipitation originated~\cite{ice_core_history}. Although
different isotopes were measured in the two separate cores, we see a close
agreement between the datasets.

The Greenland SAT proxies are plotted in green and red, based on cores from the
GRIP and NGRIP sites respectively. The green plot uses $\delta^{18}$O as a proxy
for SAT and has been converted to temperature using $T = -0.1925\delta^2
-11.88\delta - 211.4$~\cite{sat_from_d18O}. This model was fitted to ice core
data from GRIP so is likely to be more accurate than the generalised linear
model. This core has a $\delta^{18}$O record up to 250\,kya. However, due to ice
folding close to the bedrock, only the most recent 100\,kyr of the record is
reliable~\cite{last_10p_of_grip_corrupted}. We have shown the full record
here because we only wish to compare it to our solution qualitatively. The red
plot spans only 120\,kyr but is considered reliable for this duration as it is
from the separate drill site, NGRIP~\cite{last_10p_of_grip_corrupted}. The plot
is also based on $\delta^{18}$O but additionally uses the nitrogen isotope ratio
from the air bubbles trapped within the ice. Whilst air can still move around
the compacting snow, the $^{15}$N isotopes are more likely to sink than the
lighter $^{14}$N. This enrichment of the lower portion of compacting snow is
inversely related to the surrounding temperature, providing additional data for
the SAT estimate~\cite{kindler2014}.

Notably, the data from Greenland provides a higher time resolution but
significantly shorter time-span. This is due to the accumulation rates around
the drill sites~\cite{ice_core_history}. With greater precipitation, there are
more data available for the same duration. However, this reduces the time span
captured by the same depth of ice.

The increased resolution in Greenland's ice core allowed for the discovery of
the high frequency climate fluctuations known as Dansgaard–Oeschger
events~\cite{DO_1,DO_2}. The cause of these fluctuations is not fully
understood, though they appear to be local to the northern hemisphere and are
believed to relate to changes in the Atlantic ocean due to freshwater
perturbations~\cite{greenland_high_amplitude}. These oscillations occur with a
period of approximately 1.5\,kyr \cite{DO_events_period}. Since our model is a
linear combination of the significantly slower orbital parameters, these
fluctuations will not appear in our solution for SAT. This is the intended
behaviour of our model since the phenomenon is intrinsic to Earth. Instead of
reproducing the SAT data exactly, we wish to show that the SAT dynamics that
result directly from the orbital variations are sufficient to explain the
majority of the ice volume data.

Since our data only represents the most extreme of environments, we are
careful not to overgeneralise to global SAT. However, it is worth noting that
$S(t)$ might better reflect polar SAT as it holds the most significant influence
on ice volume. We therefore expect to find a reasonable qualitative agreement
between the polar SAT data and the $S(t)$ solution that best reproduces the ice
volume data.

\subsection{Fitting}
\label{sec:fitting}
To estimate each physical parameter, we use the fit $p_i$ values from our
phenomenological model along with 4 additional constraints. These relate to the
range, and average, of bulk ocean temperature and SAT. We see BWT varies between
-1$^\circ$C and 2$^\circ$C globally, whereas the mean SWT varies dependent on
location, but has an approximate range of 5$^\circ$C. Assuming these both
contribute equally to the bulk ocean temperature gives an estimated bulk ocean
temperature range of $\Delta_\mathrm{O}=4^\circ$C. The average of the SWT data
is approximately 19.4$^\circ$C whilst the averaged BWT is 0.2$^\circ$C, giving
an estimated average of $\mu_\mathrm{O}=9.8^\circ$C.

For the SAT data, we are restricted by the extreme locations of its sources,
namely Greenland and Antarctica. We assume that these reflect the qualitative
dynamics of global SAT over time but may have a differing range to the global
signal, and will certainly have a lower mean than the global average. We will
therefore employ further sources to determine the $S(t)$ equation coefficients.

Thomas estimates that SAT increased $5.8\pm1.4^\circ$C from the last glacial
maximum to pre-industrial time~\cite{surface_air_temp_range}, which was
approximately 13$^\circ$C~\cite{pre_industrial_temp}. Since these periods span
close to the minimum and maximum temperatures of the past 800\,kyr, we
use this to constrain the SAT between 7.2 and 13$^\circ$C, giving an
estimated range of $\Delta_\mathrm{S}=5.8^\circ$C and an estimated mean SAT of
$\mu_\mathrm{S}=10.1^\circ$C. We now have the necessary constraints to extract
each physical parameter from the phenomenological model coefficients, where the
equations to solve are
\begin{align}
  ac &= p_1,\\
  bd &= p_2,\\
  be &= p_3,\\
  bf &= p_4,\\
  a\alpha_O-b\alpha_S+\alpha_I &= p_5,\\
  \mathrm{Mean}[O(t)] &= \mu_\mathrm{O},\\
  \mathrm{Range}[O(t)] &= \Delta_\mathrm{O},\\
  \mathrm{Mean}[S(t)] &= \mu_\mathrm{S},\\
  \mathrm{Range}[S(t)] &= \Delta_\mathrm{S},
\end{align}
where
\begin{align}
    \mathrm{Mean}[X(t)] &= \frac{1}{800}\int_0^{800}X(t)\mathrm{d}t,\\
    \mathrm{Range}[X(t)] &= \mathrm{max}\left[X(t)\right] -
                    \mathrm{min}\left[X(t)\right],\\
    O(t) &= c\zeta_\tau[\varepsilon(t)] + \alpha_O - \alpha_O
    e^{\nicefrac{-t}{\tau}}  + O_0e^{\nicefrac{-t}{\tau}},\\
    S(t) &= d\varepsilon(t) + e\beta(t) + f\cos\left({\rho(t)}\right) +
    \alpha_S.
\end{align}
The solution to these equations produces the physical model parameters shown in
Table \ref{tab:physical_params}.
\begin{table}
  \centering
  \caption{Parameter values for our physical model.}
\label{tab:physical_params}
\begin{tabular}{c|c|c}
Parameter  & Value             & Units           \\ \hline\hline
$a$        & $2.2\times10^7$ & km$^3/^\circ$C \\ \hline
$b$        & $2.8\times10^7$ & km$^3/^\circ$C \\ \hline
$c$        & $85$           & $^\circ$C      \\ \hline
$d$        & $70$           & $^\circ$C      \\ \hline
$e$        & $56$           & $^\circ$C      \\ \hline
$f$        & $0.69$          & $^\circ$C      \\ \hline
$\alpha_I$ & $1.4\times10^8$ & km$^3$         \\ \hline
$\alpha_O$ & $7.5$           & $^\circ$C      \\ \hline
$\alpha_S$ & $-14$          & $^\circ$C      \\ \hline
$\tau$     & $15$           & kyr
\end{tabular}
\end{table}
\begin{figure}
  \vspace{-42pt}
  \input{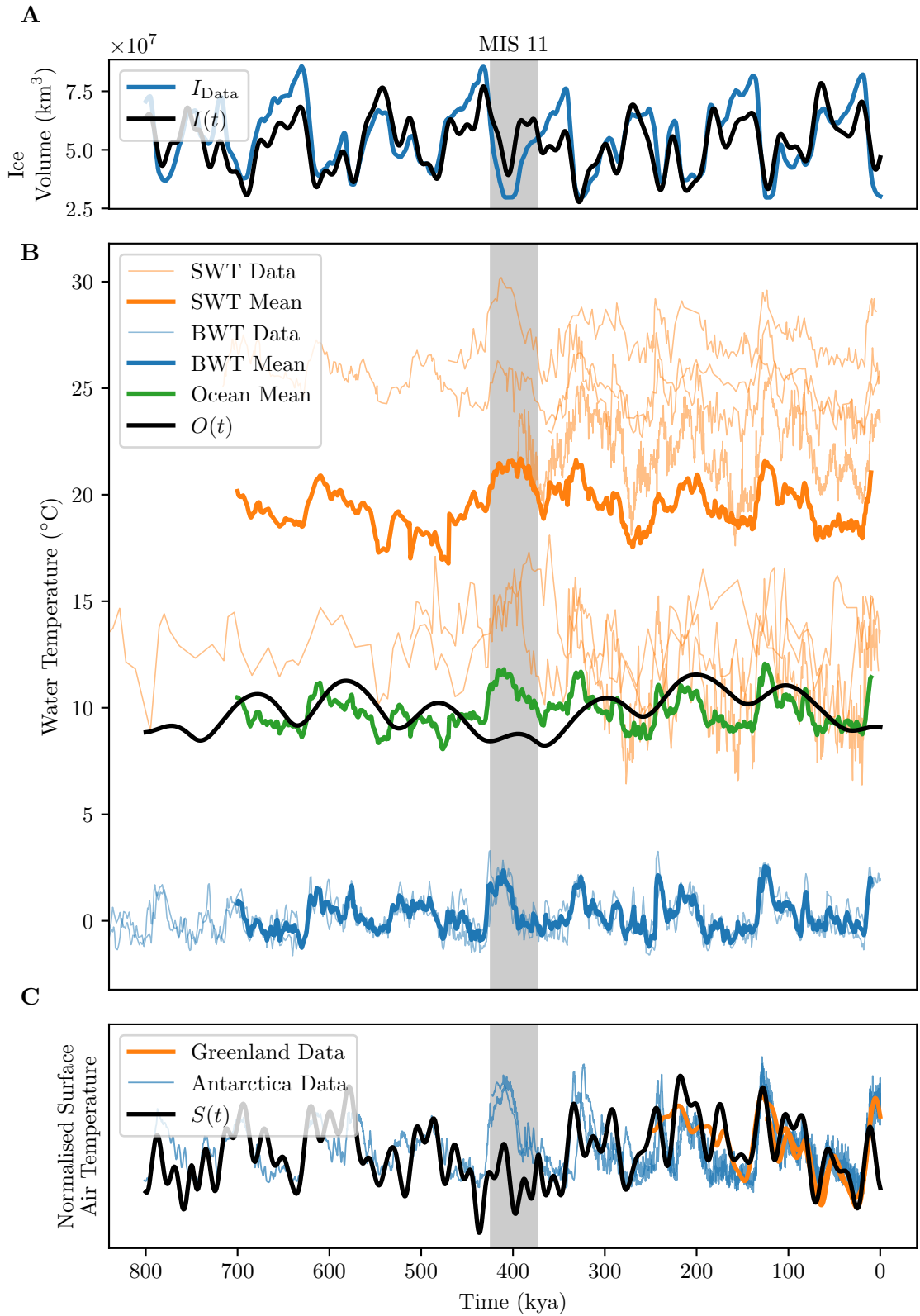}
  \caption{\textbf{A}: Our modelled ice volume compared with observational
    constraints (for details, see Appendix \ref{sec:ice_vol}). \textbf{B}: Our
    modelled bulk ocean temperature compared with the SWT and BWT data shown in
    Fig. \ref{fig:sea_and_air_temp_with_map}. The mean SWT and BWT are averaged
    to give an approximate bulk ocean temperature (green) with the ocean
    solution $O(t)$ overlaid. To avoid over dependence on one data source, the
    averages are only calculated as far as 700\,kya. \textbf{C}: Our modelled
    SAT compared with data from Fig. \ref{fig:sea_and_air_temp_with_map}. The
    data only reflects glacial air temperatures so has been rescaled to align
    with the warmer surface solution $S(t)$ to show the qualitative similarity.
    We have also only included the smoothed signal for the Greenland data as it
    is too noisy to discern the large-scale dynamics. The MIS 11 interglacial
    period is highlighted in grey.}
  \label{fig:sea_air_data_sol_comp}
\end{figure}

With these coefficients we can plot the proposed solutions for bulk ocean
temperature~$O(t)$ and SAT~$S(t)$, the weighted sum of which gives the same
solution for ice volume $I(t)$ shown in Fig. \ref{fig:fit_ice_vol}. These
solutions are all shown together, alongside their relevant data, in
Fig.~\ref{fig:sea_air_data_sol_comp}.

In comparing the solutions for $O(t)$ and $S(t)$ with their respective proxy
data, we see a similar agreement for the majority of the timespan, except around
MIS 11 as with the ice volume solution from our phenomenological model. This
artefact is discussed in Sec. \ref{sec:discussion}. Aside from MIS 11, we see
the $O(t)$ solution loosely matching the 100\,kyr oscillations that appear in
the averaged data. However, without the higher frequency obliquity and
precession as inputs to $O(t)$, it is restricted to the frequencies present in
eccentricity.

Investigating SAT data, it appears to reproduce the data well. Due to its
faster response rate, and the inclusion of both obliquity and precession, a
higher range of frequencies are captured. The quality of this fit is especially
noteworthy because we have not directly fit to the SAT data. Instead, we used
the ice volume data to determine the dynamics of $S(t)$, and then linearly
scaled the solution using SAT data. This suggests that surface temperature can
be well explained by the orbital parameters alone, and that it is a significant
factor in determining ice volume. It is also worth noting that the solutions
have similar ranges when scaled by their respective coefficients in the ice
volume equation. This suggests that their roles are of similar significance in
governing the glacial-interglacial cycles.

Now that we have determined the coefficients of our physical model, we can
convert the simplified parameters in $O(t)$ to those used in the original ocean
temperature model given by
\[
  \frac{\mathrm{d}O(t)}{\mathrm{d}t} = 9.591\gamma\varepsilon(t) + 353.8\gamma -
    mO(t) - n,%\numberthis
    \label{eq:original_ocean_temp}
\numberall
\]
where $\gamma$, $m$, and $n$ are unknown.

From fitting to the ice volume data, we attained values of $c = 85^\circ$C,
$\alpha_O = 7.5^\circ$C, and $\tau = 15$\,kyr. Using the substitutions outlined
in Sec. \ref{sec:choice_of_mechanisms}, we convert back to get
\begin{align}
    \gamma &= \frac{c}{9.591\tau} = 0.59,\\
    m &= \frac{1}{\tau} = 0.067\,\mathrm{kyr}^{-1},\\
    n &= 353.8\gamma - \frac{\alpha_O}{\tau} = 210^\circ\mathrm{C/kyr}.
\end{align}
Note that, although the constant heat loss term $n=210^\circ\mathrm{C/kyr}$
seems extreme, it is mostly cancelled out by the constant heat gain term
$353.8\gamma$, resulting in a dynamic equilibrium. With these fitted parameter
values, we can estimate the average heat loss rate of the ocean as
\begin{align}
    \mathrm{Mean}\left[mO(t)+n\right] &= m \mu_O + n,\\
    &= 0.0670\times 9.8 + 210,\\
    &= 211^\circ\mathrm{C/kyr},
\end{align}
where $\mu_O=9.8^\circ$C was estimated in Sec. \ref{sec:fitting}.

It is encouraging that once fitting the free parameters of the ocean model, we
attain a value of $\gamma=0.59$. This means almost 60\% of the extra insolation
due to eccentricity is transferred as heat to the bulk ocean, which is
physically valid. If the fitting process had produced an optimal value of
$\gamma>1$, it would suggest that some form of amplification is required to
produce the observed dynamics. By attaining a value of $\gamma=0.59$, we have
therefore shown that enough energy is present in the system for eccentricity to
explain the range of ocean temperature without the need for an amplification
term.

Additionally, the result of an almost constant heat loss rate lends support to our
original simple calculation in Sec. \ref{sec:orbital_params}, where we assumed
a constant heat loss rate and $\gamma=0.5$. Aside from these assumptions, we
used measured values to estimate a temperature rise of 6.4$^\circ$C if maximum
eccentricity was sustained for 20\,kyr. This shows good agreement with the
approximate 4$^\circ$C rise expressed in the proxy data over the same period.
Although eccentricity does not remain constant, this shows it can have a direct
and significant impact on the ocean and cryosphere without the need for
amplification.

A near constant heat loss rate is physically plausible due to the omission of
the surface ocean layer when defining $O(t)$. The surface layer exchanges heat
more easily with the atmosphere due to evaporation, whilst the deeper ocean is
more insulated. It is therefore plausible that an approximately 4$^\circ$C
change in bulk ocean temperature has little impact on its heat loss rate.

\section{Discussion}
\label{sec:discussion}
We have presented a simple phenomenological model of global ice volume, assuming
only a linear dependence on the orbital parameters. Aside from the interval around
MIS 11, this model was able to reproduce the qualitative features of the ice
volume data over the past 800\,kyr. We then proposed a physical interpretation
of the model, whereby intermediate variables representing ocean temperature and
SAT respond to the orbitally governed insolation. The weighted sum of these two
intermediate variables results in the same ice volume solution as before. The
fit coefficients were physically plausible and gave solutions that qualitatively
matched the ocean temperature and SAT data, apart from around MIS 11. This
supports the hypothesis that the majority of ice volume dynamics can be
explained by a linear model that is only driven by the orbital parameters.

\subsection{Model limitations}
For our physical model, we proposed variables $O(t)$ and $S(t)$ to represent the
slow and fast parts of the ice volume equation respectively. We emphasise that,
although these have been attributed to ocean temperature and SAT, they may also
incorporate other geophysical mechanisms. Examples of these are the CO$_2$
cycle, which is connected to ocean dynamics, and direct glacial radiation, which
would be an instantaneous function of insolation, similar to SAT. It is also
possible that, instead of modelling the global averages for each variable, we
are representing certain regions or internal components of them.

An example of this is a region known as the North Atlantic Deep Water (NADW),
which is part of the Atlantic Meridional Overturning Circulation (AMOC). This
circulation is characterised by warmer surface water that flows north towards
the NADW, where it cools, becoming denser and sinking due to convection before
flowing south. The NADW plays an important role in the oxygenation of Earth's
deep oceans and the modulation of SAT through the loss of evaporative heat
\cite{nadw_importance1}. As a result, it has been suggested that the NADW is a
significant factor in long-term climate variability \cite{nadw_importance2}.
This more abstract interpretation of $O(t)$ could also lead to a better
justification for the positive relationship between $O(t)$ and ice volume
apparent in our physical model. Another issue with $O(t)$ representing global
ocean temperature relates to the 400\,kyr period that appears in its solution.
From spectral analysis of the available ocean data, there is little evidence for
a 400\,kyr period. This could be due to the inaccurate and noisy nature of the
data, though it is also possible that more complex dynamics are influencing
ocean temperature, effectively removing the 400\,kyr period from the data.

Although our ice volume solution largely matches the local trends of the data,
it does not always match the magnitude. We present two possible explanations for
this. One possibility relates to the inherent uncertainty in the data. As
discussed in Appendix \ref{sec:ice_vol}, our chosen ice volume data has been
aggregated from a number of different sources, and the BWT component of this signal
was then removed using a model. Since the data are inherently uncertain, we
would not expect the model to fit it perfectly. A second possibility is that
Earth system feedbacks play a more significant role on ice volume at
extreme values. For example, the feedback that results from ice albedo can lead
to accelerated ice growth or retreat, depending on the direction of the change.
For moderate values of ice volume, any feedbacks were shown to be well
approximated by our linear phenomenological model. However, it appears that the
non-linear response to the orbital cycles becomes more pronounced around the
highest 25\%, and the lowest 12\%, of ice volume. As a result, the linear model
fails to match the amplitudes at these extrema in the data.

One interval where all variables fail to match even the local trend of the
data is around MIS 11. This region is both the warmest, and the longest,
interglacial over the past 800\,kyr, which could be due in part to ice sheet
instability. Hearty examined marine deposits in Bermuda and the Bahamas and
proposed that a sea level rise of approximately 20\,m above present took place
during MIS~11~\cite{mis11_ice_collapse}. Hearty suggests that a 20\,m sea level
rise could only be achieved by the complete melting of Greenland and West
Antarctica, as well as the partial melting of the far larger East Antarctica ice
sheet. This is substantiated by Christ, who uses the subglacial sediment in
an ice core from northwestern Greenland to conclude that this region was ice
free during MIS 11 \cite{greenland_ice_free_mis11}.

Ice sheets such as West Antarctica are able to flow quickly through ice-streams,
resulting in non-linear flow that is hard to model. Hollin points out that
because these ice sheets are unstable, they can succumb to surges of melting
with runaway effects~\cite{glacial_instability}. Since the dynamics of these ice
sheets are not well understood, we cannot explain why this occurred so
significantly in MIS 11 in particular. However, if this was the result of an
Earth system runaway feedback, we would not expect it to appear in our orbitally
forced solution of ice volume. We would also expect the surge in melting ice
volume to result in a reduction in global albedo and an increase in atmospheric
CO$_2$ as it is released from the ice sheets. This would then explain why ocean
temperature and SAT are significantly higher than our physical model predicts
during this period.

Another potential factor for the misalignment around MIS 11 is the Mid-Brunhes
Event, which occurred around the same time as MIS 11 and marked a global
climatic shift. The Mid-Brunhes Event is characterised by a shift to warmer
interglacial periods, along with an increase in the variance of atmospheric
CO$_2$ and CH$_4$, Antarctic temperature, and ocean
temperature~\cite{mid_brunhes_impacts}. Although the cause of this event is not
well understood, some suggest that it is due to a complex climatic response to
magnitude changes in the orbital
parameters~\cite{mid_brunhes_ecc,mid_brunhes_ecc2}.
% During MIS 11, eccentricity
% was at the minimum of its 400\,kyr period, resulting in an extended low
% amplitude. It is unclear how this could produce such a significant climatic
% shift, though a similar effect is proposed by Imbrie, only they apply the theory
% to the MPT, which occurred around 800\,kyr ago, as shown in Figure
% \ref{fig:benth_and_power_specs}. This interval coincides with the previous
% minimum in eccentricity's 400\,kyr period~\cite{imbrie2011}. Imbrie showed that
% the MPT can be reproduced using only the orbital parameters as input, proposing
% that this climatic shift was due to a non-linear response to fluctuations in
% their amplitudes. A contrasting theory of the MPT suggests that it resulted from
% a decrease in atmospheric CO$_2$, as well as the gradual removal of regolith
% from beneath the glaciers~\cite{mpt_cause}.
Because MIS 11 likely resulted from either a non-linear response to orbital
forcing, or from an Earth system process, we would not expect our linear OFPA
model to capture them.

The purpose of our model was to isolate the effect of orbital forcing on the
Earth system, and therefore it cannot produce unforced oscillations. We
recognise that intrinsic cycles on Earth likely play a role in the ice volume
dynamics, however our model demonstrates that they are not required to explain
the majority of the ice volume dynamics.

% To substantiate this, we evaluated a version of our model that
% could produce unforced oscillations. This was achieved by introducing a negative
% dependence on ice volume into the ocean temperature equation. This augmented
% model could explain approximately 7\% more variance in the data. This is a
% significant improvement, increasing the accuracy slightly more than the addition
% of our precession term, supporting the notion that intrinsic cycles play a role.

% However, aside from the intention of this paper to isolate the effect of orbital
% forcing, the inclusion of this term gives rise to a number of issues. Firstly,
% we attain a far more complex analytical solution, which makes it difficult to
% interpret the role of each term. The solution is also non-linear in multiple
% parameters. This prevents us from using the least-squares fitting approach as
% before, increasing the fitting time by orders of magnitude. This presents a
% problem when repeatedly fitting subsets of the model as is done for Figure
% \ref{fig:leave_one_out_with_diagram}.
\subsection{$Q_{65}$}
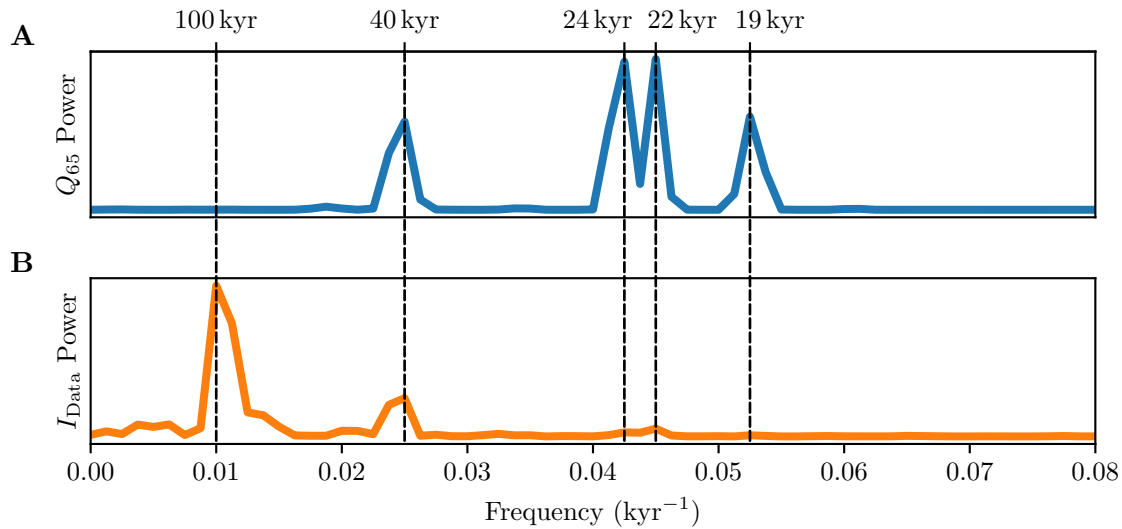
\begin{figure}
  \centering
  \input{combined_q65_benthic_power_spec.pgf}
  \caption{Power spectra for $Q_{65}$ (\textbf{A}) and the ice volume data
    (\textbf{B}) for the past 800\,kyr. Periods corresponding to peak
    frequencies are marked by dashed lines.}
  \label{fig:q65_benthic_power_spec}
\end{figure}

One of the main aims for this paper was to address the hypothesis that
eccentricity's impact on insolation is too small to drive the
glacial-interglacial cycles without amplification. Justification for this
hypothesis often leans on the $Q_{65}$ insolation measure, which is commonly
used in this area~\cite{pp04,imbrie_inertia,eg_using_insolation}. As shown in
Fig.~\ref{fig:q65_benthic_power_spec}, the 100\,kyr period relating to
eccentricity is imperceptible in the $Q_{65}$ power spectrum compared to the
frequencies relating to obliquity and precession. It is therefore logical to
hypothesise that one or more feedback mechanisms are amplifying eccentricity's
impact to produce the significant 100\,kyr peak shown in the ice volume power
spectrum.

Since $Q_{65}$ measures the insolation at a single latitude on a single day each
year, we question how well it can represent the impact of each orbital parameter
on the global insolation profile. By instead evaluating eccentricity's effect on
Earth's average annual insolation, we have shown that it is capable of producing
significant changes to the global ocean temperature without amplification.

\subsection{Physical Variables}
In Fig. \ref{fig:sea_air_data_sol_comp}, we see that $O(t)$ only weakly aligns
with the averaged ocean temperature data. This is understandable, since the
purpose of these physical variables is not to accurately reproduce their
corresponding data. Rather, their purpose is to depict the component of each
mechanism's dynamics that arise due to orbital forcing. Since our modelled ice
volume depends linearly on these two variables, we hope to have isolated
the ice volume dynamics that result from a linear response to orbital forcing.

This approach would not work if most of ocean temperature or SAT dynamics were
intrinsic or non-linear. However, we have shown that the SAT data can be well
reproduced using only the orbital parameters as inputs. The ocean temperature
data are less accurately reproduced through this method, though we were able to
demonstrate that eccentricity alone could explain the magnitude of change seen
in the data. This demonstration used measured values and the free parameter
$\gamma$, which was estimated to be $0.59$ from fitting to the ice volume data.
By attaining a fit value of $0\leq\gamma\leq 1$, the ocean temperature model
remains physically feasible, and can therefore lend support to the OFPA school
of thought.

\subsection{Sensitivity Analysis}
\begin{figure}
  \centering
  \input{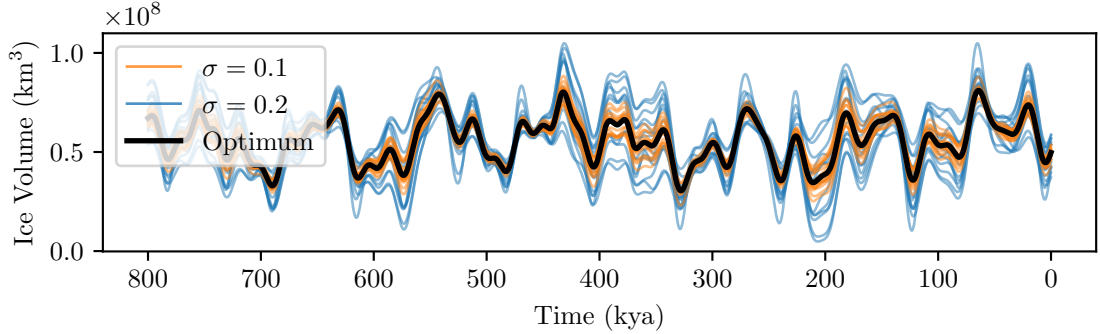}
  \caption{Set of solutions for random parameter perturbations. All parameters
    in the physical model have been perturbed each iteration accept for $\alpha_I$,
    which was chosen as a function of the other parameters each time to maintain
    the correct offset. The parameters are rescaled by coefficients chosen from
    $\mathcal{N}(1,\sigma^2)$ with simulations run for $\sigma=0.1,0.2$.}
  \label{fig:param_perturb}
\end{figure}

In order to determine the overall sensitivity of the ice volume solution on the
parameters in our physical model, we perturb them randomly and show the
resultant ice volume solution. The results of this can be seen in
Fig.~\ref{fig:param_perturb}. Here we have rescaled each of the parameters by
a different value drawn from the normal distribution $\mathcal{N}(1,\sigma^2)$.
The figure shows 15 iterations of this process for both $\sigma=0.1$ and
$\sigma=0.2$. In order to maintain the correct offset, the constant term
$\alpha_I$ was determined as a function of the other parameters.

In the $\sigma=0.1$ case, we see close agreement with the optimal fit in all 15
iterations. Only when the perturbations are drawn from $\mathcal{N}(1,0.2^2)$ do
we see a significant deviation from the optimal fit, though the qualitative
behaviour is still preserved. This suggests that the model is not highly
sensitive to any of the parameters, reducing the risk of overfitting.
Furthermore, it suggests that in the absence of anthropogenic forcing or
rare climatic shifts, the system is fairly predictable.

\subsection{Conclusion}
Here we have shown that since the MPT, global ice volume dynamics can be
approximated by a linear function of the orbital parameters. Our results show
that the data may not necessitate switching mechanisms nor unforced
oscillations, supporting the Orbital Forcing with Potential Amplification
approach.

Additionally, through estimating the annually averaged insolation reaching the
bulk ocean, we found that eccentricity is capable of producing significant
changes to the ocean and cryosphere without amplification. We propose this
important effect may have previously been overlooked due to the common use of
$Q_{65}$, which we find to be insufficient in representing the impact of each
orbital parameter on the global insolation profile.

Our phenomenological model also provides a mechanism that is able to remove
eccentricity's 400\,kyr period from the ice volume solution, in line with the
data. For our physical model, we propose this mechanism relates to the bulk
ocean temperature, which responds to eccentricity on a slower timescale than the
SAT. By taking the difference of these two variables in our ice volume equation,
we produce a signal that resembles the change in eccentricity. This signal
has significantly less power in the 400\,kyr band which results in a far better
fit to the ice volume data. To emphasise the significance of this mechanism, we
systematically refit the model with each subset of the terms in the
phenomenological model. We found the pair of lagged and instantaneous
eccentricity terms to be the greatest contributor to the model's accuracy,
suggesting that ice volume has more dependence on the change in eccentricity
that its absolute value.

Although we are modelling this Earth system as linear, we acknowledge that
non-linear dynamics are at play, and are more important at certain times.
Meltwater erosion and ice-albedo are two such feedbacks that could lead to the
accelerated change in ice volume expressed in the data. However, since the MPT,
there have only been eight 100\,kyr glacial-interglacial cycles. If we are to
unearth the mechanisms that govern this relatively brief period of ice volume
dynamics, there is value in imposing as few assumptions as possible. Since the
data are adequately reproduced by our model, we propose that ice volume dynamics
predominantly follow an approximately linear response to the orbital parameters.

\section{Acknowledgements}
L.W. gratefully acknowledges the support of the Engineering and Physical
Sciences Research Council.

T.G. gratefully acknowledges the support of the WoodNext Foundation, a component
fund administered by the Greater Houston Community Foundation.

\bibliographystyle{ieeetr}
\bibliography{refs}

\appendix
\section{Orbital Data}
\label{sec:orbital_data}
For all models in this paper, we use the orbital solutions produced by Laskar in
2004~\cite{laskar2004}. To estimate the reliability of these solutions over the
past 1\,myr, we can compare them to the previous solutions produced by Laskar in
1993~\cite{laskar1993}, as well as the solutions produced by Berger in
1999~\cite{berger1999}. Laskar's 1993 solution for eccentricity deviates from
the more recent equivalent by an average of 0.4\% over the past 1\,myr, whilst
Berger's eccentricity solution deviates by an average of 1.2\%. The obliquity
and precession deviations are significantly less in both cases. The differences
between these solutions suggests an uncertainty on the order of 1\%. This is an
acceptable level for our purposes due to the less accurate nature of the proxy
data we rely on to estimate global ice volume. To justify this, we tried fitting
our model with all three orbital solutions and found that the difference was
negligible when comparing to the ice volume data.

\section{Ice Volume Data}
\label{sec:ice_vol}
Ice volume proxy data is commonly derived by analysing isotopic concentrations
in drill cores obtained from continental ice sheets or the ocean floor. For oceanic
records, the $\delta^{18}$O value is often used. This represents the ratio of
$^{18}$O to $^{16}$O found in foraminifera shells deposited on the ocean floor.
The ratio provides insights into the proportion of $^{16}$O, which evaporates
more easily due to an effect known as Rayleigh distillation, stored within
glacial ice during a specific period. Many datasets have been produced through
this method, however there is a lack of agreement between them due to the
complexity of the reconstruction process. By aggregating multiple datasets, we
can expect the impact of anomalies and location specific behaviour to be
reduced. Lisiecki and Raymo collected $\delta^{18}$O data from 57 marine
sediment cores around the world and combined them using an automated graphic
correlation algorithm~\cite{benthic_data}. Although $\delta^{18}$O is related to
the volume of glacial ice, it is also affected by the local bottom water
temperature (BWT) in which the foraminifera lived. The ambient temperature of
the foraminifera changes the amount of $^{18}$O that is absorbed into their
shells~\cite{delta_O_to_temp}. With only one measure to represent two
quantities, researchers have used the ratio of Magnesium to Calcium in the
shells to estimate BWT. Despite this, the Mg/Ca ratio within the foraminifera
shells can also vary due to calcite dissolution, the degree of which
increases with depth~\cite{dissolution_with_depth}.

Instead of relying on additional measures, Bintanja uses the same $\delta
^{18}$O stack produced by Lisiecki and Raymo to create an internally consistent
model of mean surface air temperature, linked with ice volume and global sea
level~\cite{benthic_model_contributions}. This then produces time series for
the ice volume and BWT contributions to the $\delta^{18}$O data. The outputs
of this model were compared against other independent proxy datasets and showed
close agreement over the full timespan. We therefore opt to use Bintanja's
modelled ice volume data to fit our coefficients.

The ice volume data from Bintanja's model is expressed as a proportion of
the total benthic $\delta^{18}$O value, measured in parts-per-thousand
($\permil$). In order to evaluate our model by comparing to other
proxy data, we convert the ice proportion of $\delta^{18}$O into ice
volume. For this, we assume a linear relationship between ice volume and the
contribution to the $\delta^{18}$O data. We then use estimates for the physical
values needed to linearly transform the ice contribution into ice volume.

The current estimates of ice volume by location are approximately
$2.7{\times}10^7$\,km$^3$ in Antarctica~\cite{antarctica_ice_vol},
$2.99{\times}10^6$\,km$^3$ in Greenland~\cite{greenland_ice_vol}, and
$1.58{\times}10^5$\,km$^3$ in all other regions~\cite{other_ice_vol}.
This gives an approximate total of $3.0{\times}10^7$\,km$^3$. Lambeck
estimates that the ice volume during the last glacial maximum was approximately
$5.2{\times}10^7$\,km$^3$ greater than at present~\cite{lgm_ice_vol}. This
occurred around 21\,kya, giving us an anchor point to scale
the range of $\delta^{18}$O to ice volume. We can now convert the ice volume
component of the benthic $\delta^{18}$O data $I_\mathrm{Benth}$ into total ice
volume $I_\mathrm{Data}$ using
\[
  I_\mathrm{Data} = mI_\mathrm{Benth} + c,%\numberthis
  \label{eq:ice_data_rescale}
\numberall
\]
where
\begin{align}
    m &=
    \frac{I_\mathrm{Data}(0)-I_\mathrm{Data}(-21)}{I_\mathrm{Benth}(0)-I_\mathrm{Benth}(-21)}
    \approx \frac{3.0\times 10^7 - 8.2\times 10^7}{0.0-1.0} = 5.2\times
    10^7\,\mathrm{km}^3/\permil,\\ c &= I_\mathrm{Data}(0) -
    mI_\mathrm{Benth}(0) \approx 3.0\times 10^7 - 5.2\times 10^7\times 0.0 =
    3.0\times 10^7\,\mathrm{km}^3.
\end{align}
\end{document}

%% file: figs/combined_q65_benthic_power_spec.pgf
%% Creator: Matplotlib, PGF backend
%%
%% To include the figure in your LaTeX document, write
%%   \input{<filename>.pgf}
%%
%% Make sure the required packages are loaded in your preamble
%%   \usepackage{pgf}
%%
%% Also ensure that all the required font packages are loaded; for instance,
%% the lmodern package is sometimes necessary when using math font.
%%   \usepackage{lmodern}
%%
%% Figures using additional raster images can only be included by \input if
%% they are in the same directory as the main LaTeX file. For loading figures
%% from other directories you can use the `import` package
%%   \usepackage{import}
%%
%% and then include the figures with
%%   \import{<path to file>}{<filename>.pgf}
%%
%% Matplotlib used the following preamble
%%   \usepackage{wasysym}\usepackage{nicefrac}
%%   \makeatletter\@ifpackageloaded{underscore}{}{\usepackage[strings]{underscore}}\makeatother
%%
\begingroup%
\makeatletter%
\begin{pgfpicture}%
\pgfpathrectangle{\pgfpointorigin}{\pgfqpoint{6.000000in}{3.000000in}}%
\pgfusepath{use as bounding box, clip}%
\begin{pgfscope}%
\pgfsetbuttcap%
\pgfsetmiterjoin%
\definecolor{currentfill}{rgb}{1.000000,1.000000,1.000000}%
\pgfsetfillcolor{currentfill}%
\pgfsetlinewidth{0.000000pt}%
\definecolor{currentstroke}{rgb}{1.000000,1.000000,1.000000}%
\pgfsetstrokecolor{currentstroke}%
\pgfsetdash{}{0pt}%
\pgfpathmoveto{\pgfqpoint{0.000000in}{0.000000in}}%
\pgfpathlineto{\pgfqpoint{6.000000in}{0.000000in}}%
\pgfpathlineto{\pgfqpoint{6.000000in}{3.000000in}}%
\pgfpathlineto{\pgfqpoint{0.000000in}{3.000000in}}%
\pgfpathlineto{\pgfqpoint{0.000000in}{0.000000in}}%
\pgfpathclose%
\pgfusepath{fill}%
\end{pgfscope}%
\begin{pgfscope}%
\pgfsetbuttcap%
\pgfsetmiterjoin%
\definecolor{currentfill}{rgb}{1.000000,1.000000,1.000000}%
\pgfsetfillcolor{currentfill}%
\pgfsetlinewidth{0.000000pt}%
\definecolor{currentstroke}{rgb}{0.000000,0.000000,0.000000}%
\pgfsetstrokecolor{currentstroke}%
\pgfsetstrokeopacity{0.000000}%
\pgfsetdash{}{0pt}%
\pgfpathmoveto{\pgfqpoint{0.522000in}{1.764139in}}%
\pgfpathlineto{\pgfqpoint{5.726543in}{1.764139in}}%
\pgfpathlineto{\pgfqpoint{5.726543in}{2.629321in}}%
\pgfpathlineto{\pgfqpoint{0.522000in}{2.629321in}}%
\pgfpathlineto{\pgfqpoint{0.522000in}{1.764139in}}%
\pgfpathclose%
\pgfusepath{fill}%
\end{pgfscope}%
\begin{pgfscope}%
\pgfsetbuttcap%
\pgfsetmiterjoin%
\definecolor{currentfill}{rgb}{1.000000,1.000000,1.000000}%
\pgfsetfillcolor{currentfill}%
\pgfsetlinewidth{0.000000pt}%
\definecolor{currentstroke}{rgb}{0.000000,0.000000,0.000000}%
\pgfsetstrokecolor{currentstroke}%
\pgfsetstrokeopacity{0.000000}%
\pgfsetdash{}{0pt}%
\pgfpathmoveto{\pgfqpoint{0.522000in}{2.629321in}}%
\pgfpathlineto{\pgfqpoint{5.726543in}{2.629321in}}%
\pgfpathlineto{\pgfqpoint{5.726543in}{2.629321in}}%
\pgfpathlineto{\pgfqpoint{0.522000in}{2.629321in}}%
\pgfpathlineto{\pgfqpoint{0.522000in}{2.629321in}}%
\pgfpathclose%
\pgfusepath{fill}%
\end{pgfscope}%
\begin{pgfscope}%
\pgfsetbuttcap%
\pgfsetroundjoin%
\definecolor{currentfill}{rgb}{0.000000,0.000000,0.000000}%
\pgfsetfillcolor{currentfill}%
\pgfsetlinewidth{0.803000pt}%
\definecolor{currentstroke}{rgb}{0.000000,0.000000,0.000000}%
\pgfsetstrokecolor{currentstroke}%
\pgfsetdash{}{0pt}%
\pgfsys@defobject{currentmarker}{\pgfqpoint{0.000000in}{0.000000in}}{\pgfqpoint{0.000000in}{0.048611in}}{%
\pgfpathmoveto{\pgfqpoint{0.000000in}{0.000000in}}%
\pgfpathlineto{\pgfqpoint{0.000000in}{0.048611in}}%
\pgfusepath{stroke,fill}%
}%
\begin{pgfscope}%
\pgfsys@transformshift{3.937439in}{2.629321in}%
\pgfsys@useobject{currentmarker}{}%
\end{pgfscope}%
\end{pgfscope}%
\begin{pgfscope}%
\definecolor{textcolor}{rgb}{0.000000,0.000000,0.000000}%
\pgfsetstrokecolor{textcolor}%
\pgfsetfillcolor{textcolor}%
\pgftext[x=3.937439in,y=2.726543in,,bottom]{\color{textcolor}\rmfamily\fontsize{10.000000}{12.000000}\selectfont \hspace{15pt}19\,kyr}%
\end{pgfscope}%
\begin{pgfscope}%
\pgfsetbuttcap%
\pgfsetroundjoin%
\definecolor{currentfill}{rgb}{0.000000,0.000000,0.000000}%
\pgfsetfillcolor{currentfill}%
\pgfsetlinewidth{0.803000pt}%
\definecolor{currentstroke}{rgb}{0.000000,0.000000,0.000000}%
\pgfsetstrokecolor{currentstroke}%
\pgfsetdash{}{0pt}%
\pgfsys@defobject{currentmarker}{\pgfqpoint{0.000000in}{0.000000in}}{\pgfqpoint{0.000000in}{0.048611in}}{%
\pgfpathmoveto{\pgfqpoint{0.000000in}{0.000000in}}%
\pgfpathlineto{\pgfqpoint{0.000000in}{0.048611in}}%
\pgfusepath{stroke,fill}%
}%
\begin{pgfscope}%
\pgfsys@transformshift{2.148399in}{2.629321in}%
\pgfsys@useobject{currentmarker}{}%
\end{pgfscope}%
\end{pgfscope}%
\begin{pgfscope}%
\definecolor{textcolor}{rgb}{0.000000,0.000000,0.000000}%
\pgfsetstrokecolor{textcolor}%
\pgfsetfillcolor{textcolor}%
\pgftext[x=2.148399in,y=2.726543in,,bottom]{\color{textcolor}\rmfamily\fontsize{10.000000}{12.000000}\selectfont 40\,kyr}%
\end{pgfscope}%
\begin{pgfscope}%
\pgfsetbuttcap%
\pgfsetroundjoin%
\definecolor{currentfill}{rgb}{0.000000,0.000000,0.000000}%
\pgfsetfillcolor{currentfill}%
\pgfsetlinewidth{0.803000pt}%
\definecolor{currentstroke}{rgb}{0.000000,0.000000,0.000000}%
\pgfsetstrokecolor{currentstroke}%
\pgfsetdash{}{0pt}%
\pgfsys@defobject{currentmarker}{\pgfqpoint{0.000000in}{0.000000in}}{\pgfqpoint{0.000000in}{0.048611in}}{%
\pgfpathmoveto{\pgfqpoint{0.000000in}{0.000000in}}%
\pgfpathlineto{\pgfqpoint{0.000000in}{0.048611in}}%
\pgfusepath{stroke,fill}%
}%
\begin{pgfscope}%
\pgfsys@transformshift{3.286879in}{2.629321in}%
\pgfsys@useobject{currentmarker}{}%
\end{pgfscope}%
\end{pgfscope}%
\begin{pgfscope}%
\definecolor{textcolor}{rgb}{0.000000,0.000000,0.000000}%
\pgfsetstrokecolor{textcolor}%
\pgfsetfillcolor{textcolor}%
\pgftext[x=3.286879in,y=2.726543in,,bottom]{\color{textcolor}\rmfamily\fontsize{10.000000}{12.000000}\selectfont 24\,kyr\hspace{20pt}}%
\end{pgfscope}%
\begin{pgfscope}%
\pgfsetbuttcap%
\pgfsetroundjoin%
\definecolor{currentfill}{rgb}{0.000000,0.000000,0.000000}%
\pgfsetfillcolor{currentfill}%
\pgfsetlinewidth{0.803000pt}%
\definecolor{currentstroke}{rgb}{0.000000,0.000000,0.000000}%
\pgfsetstrokecolor{currentstroke}%
\pgfsetdash{}{0pt}%
\pgfsys@defobject{currentmarker}{\pgfqpoint{0.000000in}{0.000000in}}{\pgfqpoint{0.000000in}{0.048611in}}{%
\pgfpathmoveto{\pgfqpoint{0.000000in}{0.000000in}}%
\pgfpathlineto{\pgfqpoint{0.000000in}{0.048611in}}%
\pgfusepath{stroke,fill}%
}%
\begin{pgfscope}%
\pgfsys@transformshift{3.449519in}{2.629321in}%
\pgfsys@useobject{currentmarker}{}%
\end{pgfscope}%
\end{pgfscope}%
\begin{pgfscope}%
\definecolor{textcolor}{rgb}{0.000000,0.000000,0.000000}%
\pgfsetstrokecolor{textcolor}%
\pgfsetfillcolor{textcolor}%
\pgftext[x=3.449519in,y=2.726543in,,bottom]{\color{textcolor}\rmfamily\fontsize{10.000000}{12.000000}\selectfont \hspace{20pt}22\,kyr}%
\end{pgfscope}%
\begin{pgfscope}%
\pgfsetbuttcap%
\pgfsetroundjoin%
\definecolor{currentfill}{rgb}{0.000000,0.000000,0.000000}%
\pgfsetfillcolor{currentfill}%
\pgfsetlinewidth{0.803000pt}%
\definecolor{currentstroke}{rgb}{0.000000,0.000000,0.000000}%
\pgfsetstrokecolor{currentstroke}%
\pgfsetdash{}{0pt}%
\pgfsys@defobject{currentmarker}{\pgfqpoint{0.000000in}{0.000000in}}{\pgfqpoint{0.000000in}{0.048611in}}{%
\pgfpathmoveto{\pgfqpoint{0.000000in}{0.000000in}}%
\pgfpathlineto{\pgfqpoint{0.000000in}{0.048611in}}%
\pgfusepath{stroke,fill}%
}%
\begin{pgfscope}%
\pgfsys@transformshift{1.172560in}{2.629321in}%
\pgfsys@useobject{currentmarker}{}%
\end{pgfscope}%
\end{pgfscope}%
\begin{pgfscope}%
\definecolor{textcolor}{rgb}{0.000000,0.000000,0.000000}%
\pgfsetstrokecolor{textcolor}%
\pgfsetfillcolor{textcolor}%
\pgftext[x=1.172560in,y=2.726543in,,bottom]{\color{textcolor}\rmfamily\fontsize{10.000000}{12.000000}\selectfont 100\,kyr}%
\end{pgfscope}%
\begin{pgfscope}%
\pgfsetrectcap%
\pgfsetmiterjoin%
\pgfsetlinewidth{0.803000pt}%
\definecolor{currentstroke}{rgb}{0.000000,0.000000,0.000000}%
\pgfsetstrokecolor{currentstroke}%
\pgfsetdash{}{0pt}%
\pgfpathmoveto{\pgfqpoint{0.522000in}{2.629321in}}%
\pgfpathlineto{\pgfqpoint{5.726543in}{2.629321in}}%
\pgfusepath{stroke}%
\end{pgfscope}%
\begin{pgfscope}%
\definecolor{textcolor}{rgb}{0.000000,0.000000,0.000000}%
\pgfsetstrokecolor{textcolor}%
\pgfsetfillcolor{textcolor}%
\pgftext[x=0.466444in,y=2.196730in,,bottom,rotate=90.000000]{\color{textcolor}\rmfamily\fontsize{10.000000}{12.000000}\selectfont \(\displaystyle Q_{65}\) Power}%
\end{pgfscope}%
\begin{pgfscope}%
\pgfpathrectangle{\pgfqpoint{0.522000in}{1.764139in}}{\pgfqpoint{5.204543in}{0.865182in}}%
\pgfusepath{clip}%
\pgfsetrectcap%
\pgfsetroundjoin%
\pgfsetlinewidth{3.011250pt}%
\definecolor{currentstroke}{rgb}{0.121569,0.466667,0.705882}%
\pgfsetstrokecolor{currentstroke}%
\pgfsetdash{}{0pt}%
\pgfpathmoveto{\pgfqpoint{0.522000in}{1.803573in}}%
\pgfpathlineto{\pgfqpoint{0.603320in}{1.804566in}}%
\pgfpathlineto{\pgfqpoint{0.684640in}{1.805189in}}%
\pgfpathlineto{\pgfqpoint{0.765960in}{1.803805in}}%
\pgfpathlineto{\pgfqpoint{0.847280in}{1.803511in}}%
\pgfpathlineto{\pgfqpoint{0.928600in}{1.803492in}}%
\pgfpathlineto{\pgfqpoint{1.009920in}{1.804213in}}%
\pgfpathlineto{\pgfqpoint{1.091240in}{1.803910in}}%
\pgfpathlineto{\pgfqpoint{1.172560in}{1.804044in}}%
\pgfpathlineto{\pgfqpoint{1.253880in}{1.804247in}}%
\pgfpathlineto{\pgfqpoint{1.335200in}{1.803490in}}%
\pgfpathlineto{\pgfqpoint{1.416520in}{1.803469in}}%
\pgfpathlineto{\pgfqpoint{1.497840in}{1.803472in}}%
\pgfpathlineto{\pgfqpoint{1.579160in}{1.803804in}}%
\pgfpathlineto{\pgfqpoint{1.660480in}{1.809029in}}%
\pgfpathlineto{\pgfqpoint{1.741799in}{1.819408in}}%
\pgfpathlineto{\pgfqpoint{1.823119in}{1.810143in}}%
\pgfpathlineto{\pgfqpoint{1.904439in}{1.803747in}}%
\pgfpathlineto{\pgfqpoint{1.985759in}{1.809526in}}%
\pgfpathlineto{\pgfqpoint{2.067079in}{2.099844in}}%
\pgfpathlineto{\pgfqpoint{2.148399in}{2.263303in}}%
\pgfpathlineto{\pgfqpoint{2.229719in}{1.857129in}}%
\pgfpathlineto{\pgfqpoint{2.311039in}{1.804693in}}%
\pgfpathlineto{\pgfqpoint{2.392359in}{1.803790in}}%
\pgfpathlineto{\pgfqpoint{2.473679in}{1.803502in}}%
\pgfpathlineto{\pgfqpoint{2.554999in}{1.803466in}}%
\pgfpathlineto{\pgfqpoint{2.636319in}{1.804666in}}%
\pgfpathlineto{\pgfqpoint{2.717639in}{1.810640in}}%
\pgfpathlineto{\pgfqpoint{2.798959in}{1.809727in}}%
\pgfpathlineto{\pgfqpoint{2.880279in}{1.803730in}}%
\pgfpathlineto{\pgfqpoint{2.961599in}{1.803506in}}%
\pgfpathlineto{\pgfqpoint{3.042919in}{1.803523in}}%
\pgfpathlineto{\pgfqpoint{3.124239in}{1.805154in}}%
\pgfpathlineto{\pgfqpoint{3.205559in}{2.227336in}}%
\pgfpathlineto{\pgfqpoint{3.286879in}{2.576193in}}%
\pgfpathlineto{\pgfqpoint{3.368199in}{1.938138in}}%
\pgfpathlineto{\pgfqpoint{3.449519in}{2.589995in}}%
\pgfpathlineto{\pgfqpoint{3.530839in}{1.871553in}}%
\pgfpathlineto{\pgfqpoint{3.612159in}{1.803965in}}%
\pgfpathlineto{\pgfqpoint{3.693479in}{1.803476in}}%
\pgfpathlineto{\pgfqpoint{3.774799in}{1.803496in}}%
\pgfpathlineto{\pgfqpoint{3.856119in}{1.885609in}}%
\pgfpathlineto{\pgfqpoint{3.937439in}{2.289264in}}%
\pgfpathlineto{\pgfqpoint{4.018759in}{1.999943in}}%
\pgfpathlineto{\pgfqpoint{4.100078in}{1.804685in}}%
\pgfpathlineto{\pgfqpoint{4.181398in}{1.803554in}}%
\pgfpathlineto{\pgfqpoint{4.262718in}{1.803481in}}%
\pgfpathlineto{\pgfqpoint{4.344038in}{1.803538in}}%
\pgfpathlineto{\pgfqpoint{4.425358in}{1.807326in}}%
\pgfpathlineto{\pgfqpoint{4.506678in}{1.808175in}}%
\pgfpathlineto{\pgfqpoint{4.587998in}{1.803657in}}%
\pgfpathlineto{\pgfqpoint{4.669318in}{1.803486in}}%
\pgfpathlineto{\pgfqpoint{4.750638in}{1.803469in}}%
\pgfpathlineto{\pgfqpoint{4.831958in}{1.803504in}}%
\pgfpathlineto{\pgfqpoint{4.913278in}{1.803489in}}%
\pgfpathlineto{\pgfqpoint{4.994598in}{1.803500in}}%
\pgfpathlineto{\pgfqpoint{5.075918in}{1.803491in}}%
\pgfpathlineto{\pgfqpoint{5.157238in}{1.803468in}}%
\pgfpathlineto{\pgfqpoint{5.238558in}{1.803466in}}%
\pgfpathlineto{\pgfqpoint{5.319878in}{1.803466in}}%
\pgfpathlineto{\pgfqpoint{5.401198in}{1.803466in}}%
\pgfpathlineto{\pgfqpoint{5.482518in}{1.803475in}}%
\pgfpathlineto{\pgfqpoint{5.563838in}{1.803495in}}%
\pgfpathlineto{\pgfqpoint{5.645158in}{1.803479in}}%
\pgfpathlineto{\pgfqpoint{5.726478in}{1.803467in}}%
\pgfpathlineto{\pgfqpoint{5.736543in}{1.803467in}}%
\pgfusepath{stroke}%
\end{pgfscope}%
\begin{pgfscope}%
\pgfsetrectcap%
\pgfsetmiterjoin%
\pgfsetlinewidth{0.803000pt}%
\definecolor{currentstroke}{rgb}{0.000000,0.000000,0.000000}%
\pgfsetstrokecolor{currentstroke}%
\pgfsetdash{}{0pt}%
\pgfpathmoveto{\pgfqpoint{0.522000in}{1.764139in}}%
\pgfpathlineto{\pgfqpoint{0.522000in}{2.629321in}}%
\pgfusepath{stroke}%
\end{pgfscope}%
\begin{pgfscope}%
\pgfsetrectcap%
\pgfsetmiterjoin%
\pgfsetlinewidth{0.803000pt}%
\definecolor{currentstroke}{rgb}{0.000000,0.000000,0.000000}%
\pgfsetstrokecolor{currentstroke}%
\pgfsetdash{}{0pt}%
\pgfpathmoveto{\pgfqpoint{5.726543in}{1.764139in}}%
\pgfpathlineto{\pgfqpoint{5.726543in}{2.629321in}}%
\pgfusepath{stroke}%
\end{pgfscope}%
\begin{pgfscope}%
\pgfsetrectcap%
\pgfsetmiterjoin%
\pgfsetlinewidth{0.803000pt}%
\definecolor{currentstroke}{rgb}{0.000000,0.000000,0.000000}%
\pgfsetstrokecolor{currentstroke}%
\pgfsetdash{}{0pt}%
\pgfpathmoveto{\pgfqpoint{0.522000in}{1.764139in}}%
\pgfpathlineto{\pgfqpoint{5.726543in}{1.764139in}}%
\pgfusepath{stroke}%
\end{pgfscope}%
\begin{pgfscope}%
\pgfsetrectcap%
\pgfsetmiterjoin%
\pgfsetlinewidth{0.803000pt}%
\definecolor{currentstroke}{rgb}{0.000000,0.000000,0.000000}%
\pgfsetstrokecolor{currentstroke}%
\pgfsetdash{}{0pt}%
\pgfpathmoveto{\pgfqpoint{0.522000in}{2.629321in}}%
\pgfpathlineto{\pgfqpoint{5.726543in}{2.629321in}}%
\pgfusepath{stroke}%
\end{pgfscope}%
\begin{pgfscope}%
\definecolor{textcolor}{rgb}{0.000000,0.000000,0.000000}%
\pgfsetstrokecolor{textcolor}%
\pgfsetfillcolor{textcolor}%
\pgftext[x=0.105637in,y=2.767750in,left,top]{\color{textcolor}\rmfamily\fontsize{11.000000}{13.200000}\bfseries\selectfont A}%
\end{pgfscope}%
\begin{pgfscope}%
\pgfsetbuttcap%
\pgfsetmiterjoin%
\definecolor{currentfill}{rgb}{1.000000,1.000000,1.000000}%
\pgfsetfillcolor{currentfill}%
\pgfsetlinewidth{0.000000pt}%
\definecolor{currentstroke}{rgb}{0.000000,0.000000,0.000000}%
\pgfsetstrokecolor{currentstroke}%
\pgfsetstrokeopacity{0.000000}%
\pgfsetdash{}{0pt}%
\pgfpathmoveto{\pgfqpoint{0.522000in}{0.580957in}}%
\pgfpathlineto{\pgfqpoint{5.726543in}{0.580957in}}%
\pgfpathlineto{\pgfqpoint{5.726543in}{1.446139in}}%
\pgfpathlineto{\pgfqpoint{0.522000in}{1.446139in}}%
\pgfpathlineto{\pgfqpoint{0.522000in}{0.580957in}}%
\pgfpathclose%
\pgfusepath{fill}%
\end{pgfscope}%
\begin{pgfscope}%
\pgfsetbuttcap%
\pgfsetroundjoin%
\definecolor{currentfill}{rgb}{0.000000,0.000000,0.000000}%
\pgfsetfillcolor{currentfill}%
\pgfsetlinewidth{0.803000pt}%
\definecolor{currentstroke}{rgb}{0.000000,0.000000,0.000000}%
\pgfsetstrokecolor{currentstroke}%
\pgfsetdash{}{0pt}%
\pgfsys@defobject{currentmarker}{\pgfqpoint{0.000000in}{-0.048611in}}{\pgfqpoint{0.000000in}{0.000000in}}{%
\pgfpathmoveto{\pgfqpoint{0.000000in}{0.000000in}}%
\pgfpathlineto{\pgfqpoint{0.000000in}{-0.048611in}}%
\pgfusepath{stroke,fill}%
}%
\begin{pgfscope}%
\pgfsys@transformshift{0.522000in}{0.580957in}%
\pgfsys@useobject{currentmarker}{}%
\end{pgfscope}%
\end{pgfscope}%
\begin{pgfscope}%
\definecolor{textcolor}{rgb}{0.000000,0.000000,0.000000}%
\pgfsetstrokecolor{textcolor}%
\pgfsetfillcolor{textcolor}%
\pgftext[x=0.522000in,y=0.483735in,,top]{\color{textcolor}\rmfamily\fontsize{10.000000}{12.000000}\selectfont \(\displaystyle {0.00}\)}%
\end{pgfscope}%
\begin{pgfscope}%
\pgfsetbuttcap%
\pgfsetroundjoin%
\definecolor{currentfill}{rgb}{0.000000,0.000000,0.000000}%
\pgfsetfillcolor{currentfill}%
\pgfsetlinewidth{0.803000pt}%
\definecolor{currentstroke}{rgb}{0.000000,0.000000,0.000000}%
\pgfsetstrokecolor{currentstroke}%
\pgfsetdash{}{0pt}%
\pgfsys@defobject{currentmarker}{\pgfqpoint{0.000000in}{-0.048611in}}{\pgfqpoint{0.000000in}{0.000000in}}{%
\pgfpathmoveto{\pgfqpoint{0.000000in}{0.000000in}}%
\pgfpathlineto{\pgfqpoint{0.000000in}{-0.048611in}}%
\pgfusepath{stroke,fill}%
}%
\begin{pgfscope}%
\pgfsys@transformshift{1.172568in}{0.580957in}%
\pgfsys@useobject{currentmarker}{}%
\end{pgfscope}%
\end{pgfscope}%
\begin{pgfscope}%
\definecolor{textcolor}{rgb}{0.000000,0.000000,0.000000}%
\pgfsetstrokecolor{textcolor}%
\pgfsetfillcolor{textcolor}%
\pgftext[x=1.172568in,y=0.483735in,,top]{\color{textcolor}\rmfamily\fontsize{10.000000}{12.000000}\selectfont \(\displaystyle {0.01}\)}%
\end{pgfscope}%
\begin{pgfscope}%
\pgfsetbuttcap%
\pgfsetroundjoin%
\definecolor{currentfill}{rgb}{0.000000,0.000000,0.000000}%
\pgfsetfillcolor{currentfill}%
\pgfsetlinewidth{0.803000pt}%
\definecolor{currentstroke}{rgb}{0.000000,0.000000,0.000000}%
\pgfsetstrokecolor{currentstroke}%
\pgfsetdash{}{0pt}%
\pgfsys@defobject{currentmarker}{\pgfqpoint{0.000000in}{-0.048611in}}{\pgfqpoint{0.000000in}{0.000000in}}{%
\pgfpathmoveto{\pgfqpoint{0.000000in}{0.000000in}}%
\pgfpathlineto{\pgfqpoint{0.000000in}{-0.048611in}}%
\pgfusepath{stroke,fill}%
}%
\begin{pgfscope}%
\pgfsys@transformshift{1.823136in}{0.580957in}%
\pgfsys@useobject{currentmarker}{}%
\end{pgfscope}%
\end{pgfscope}%
\begin{pgfscope}%
\definecolor{textcolor}{rgb}{0.000000,0.000000,0.000000}%
\pgfsetstrokecolor{textcolor}%
\pgfsetfillcolor{textcolor}%
\pgftext[x=1.823136in,y=0.483735in,,top]{\color{textcolor}\rmfamily\fontsize{10.000000}{12.000000}\selectfont \(\displaystyle {0.02}\)}%
\end{pgfscope}%
\begin{pgfscope}%
\pgfsetbuttcap%
\pgfsetroundjoin%
\definecolor{currentfill}{rgb}{0.000000,0.000000,0.000000}%
\pgfsetfillcolor{currentfill}%
\pgfsetlinewidth{0.803000pt}%
\definecolor{currentstroke}{rgb}{0.000000,0.000000,0.000000}%
\pgfsetstrokecolor{currentstroke}%
\pgfsetdash{}{0pt}%
\pgfsys@defobject{currentmarker}{\pgfqpoint{0.000000in}{-0.048611in}}{\pgfqpoint{0.000000in}{0.000000in}}{%
\pgfpathmoveto{\pgfqpoint{0.000000in}{0.000000in}}%
\pgfpathlineto{\pgfqpoint{0.000000in}{-0.048611in}}%
\pgfusepath{stroke,fill}%
}%
\begin{pgfscope}%
\pgfsys@transformshift{2.473704in}{0.580957in}%
\pgfsys@useobject{currentmarker}{}%
\end{pgfscope}%
\end{pgfscope}%
\begin{pgfscope}%
\definecolor{textcolor}{rgb}{0.000000,0.000000,0.000000}%
\pgfsetstrokecolor{textcolor}%
\pgfsetfillcolor{textcolor}%
\pgftext[x=2.473704in,y=0.483735in,,top]{\color{textcolor}\rmfamily\fontsize{10.000000}{12.000000}\selectfont \(\displaystyle {0.03}\)}%
\end{pgfscope}%
\begin{pgfscope}%
\pgfsetbuttcap%
\pgfsetroundjoin%
\definecolor{currentfill}{rgb}{0.000000,0.000000,0.000000}%
\pgfsetfillcolor{currentfill}%
\pgfsetlinewidth{0.803000pt}%
\definecolor{currentstroke}{rgb}{0.000000,0.000000,0.000000}%
\pgfsetstrokecolor{currentstroke}%
\pgfsetdash{}{0pt}%
\pgfsys@defobject{currentmarker}{\pgfqpoint{0.000000in}{-0.048611in}}{\pgfqpoint{0.000000in}{0.000000in}}{%
\pgfpathmoveto{\pgfqpoint{0.000000in}{0.000000in}}%
\pgfpathlineto{\pgfqpoint{0.000000in}{-0.048611in}}%
\pgfusepath{stroke,fill}%
}%
\begin{pgfscope}%
\pgfsys@transformshift{3.124271in}{0.580957in}%
\pgfsys@useobject{currentmarker}{}%
\end{pgfscope}%
\end{pgfscope}%
\begin{pgfscope}%
\definecolor{textcolor}{rgb}{0.000000,0.000000,0.000000}%
\pgfsetstrokecolor{textcolor}%
\pgfsetfillcolor{textcolor}%
\pgftext[x=3.124271in,y=0.483735in,,top]{\color{textcolor}\rmfamily\fontsize{10.000000}{12.000000}\selectfont \(\displaystyle {0.04}\)}%
\end{pgfscope}%
\begin{pgfscope}%
\pgfsetbuttcap%
\pgfsetroundjoin%
\definecolor{currentfill}{rgb}{0.000000,0.000000,0.000000}%
\pgfsetfillcolor{currentfill}%
\pgfsetlinewidth{0.803000pt}%
\definecolor{currentstroke}{rgb}{0.000000,0.000000,0.000000}%
\pgfsetstrokecolor{currentstroke}%
\pgfsetdash{}{0pt}%
\pgfsys@defobject{currentmarker}{\pgfqpoint{0.000000in}{-0.048611in}}{\pgfqpoint{0.000000in}{0.000000in}}{%
\pgfpathmoveto{\pgfqpoint{0.000000in}{0.000000in}}%
\pgfpathlineto{\pgfqpoint{0.000000in}{-0.048611in}}%
\pgfusepath{stroke,fill}%
}%
\begin{pgfscope}%
\pgfsys@transformshift{3.774839in}{0.580957in}%
\pgfsys@useobject{currentmarker}{}%
\end{pgfscope}%
\end{pgfscope}%
\begin{pgfscope}%
\definecolor{textcolor}{rgb}{0.000000,0.000000,0.000000}%
\pgfsetstrokecolor{textcolor}%
\pgfsetfillcolor{textcolor}%
\pgftext[x=3.774839in,y=0.483735in,,top]{\color{textcolor}\rmfamily\fontsize{10.000000}{12.000000}\selectfont \(\displaystyle {0.05}\)}%
\end{pgfscope}%
\begin{pgfscope}%
\pgfsetbuttcap%
\pgfsetroundjoin%
\definecolor{currentfill}{rgb}{0.000000,0.000000,0.000000}%
\pgfsetfillcolor{currentfill}%
\pgfsetlinewidth{0.803000pt}%
\definecolor{currentstroke}{rgb}{0.000000,0.000000,0.000000}%
\pgfsetstrokecolor{currentstroke}%
\pgfsetdash{}{0pt}%
\pgfsys@defobject{currentmarker}{\pgfqpoint{0.000000in}{-0.048611in}}{\pgfqpoint{0.000000in}{0.000000in}}{%
\pgfpathmoveto{\pgfqpoint{0.000000in}{0.000000in}}%
\pgfpathlineto{\pgfqpoint{0.000000in}{-0.048611in}}%
\pgfusepath{stroke,fill}%
}%
\begin{pgfscope}%
\pgfsys@transformshift{4.425407in}{0.580957in}%
\pgfsys@useobject{currentmarker}{}%
\end{pgfscope}%
\end{pgfscope}%
\begin{pgfscope}%
\definecolor{textcolor}{rgb}{0.000000,0.000000,0.000000}%
\pgfsetstrokecolor{textcolor}%
\pgfsetfillcolor{textcolor}%
\pgftext[x=4.425407in,y=0.483735in,,top]{\color{textcolor}\rmfamily\fontsize{10.000000}{12.000000}\selectfont \(\displaystyle {0.06}\)}%
\end{pgfscope}%
\begin{pgfscope}%
\pgfsetbuttcap%
\pgfsetroundjoin%
\definecolor{currentfill}{rgb}{0.000000,0.000000,0.000000}%
\pgfsetfillcolor{currentfill}%
\pgfsetlinewidth{0.803000pt}%
\definecolor{currentstroke}{rgb}{0.000000,0.000000,0.000000}%
\pgfsetstrokecolor{currentstroke}%
\pgfsetdash{}{0pt}%
\pgfsys@defobject{currentmarker}{\pgfqpoint{0.000000in}{-0.048611in}}{\pgfqpoint{0.000000in}{0.000000in}}{%
\pgfpathmoveto{\pgfqpoint{0.000000in}{0.000000in}}%
\pgfpathlineto{\pgfqpoint{0.000000in}{-0.048611in}}%
\pgfusepath{stroke,fill}%
}%
\begin{pgfscope}%
\pgfsys@transformshift{5.075975in}{0.580957in}%
\pgfsys@useobject{currentmarker}{}%
\end{pgfscope}%
\end{pgfscope}%
\begin{pgfscope}%
\definecolor{textcolor}{rgb}{0.000000,0.000000,0.000000}%
\pgfsetstrokecolor{textcolor}%
\pgfsetfillcolor{textcolor}%
\pgftext[x=5.075975in,y=0.483735in,,top]{\color{textcolor}\rmfamily\fontsize{10.000000}{12.000000}\selectfont \(\displaystyle {0.07}\)}%
\end{pgfscope}%
\begin{pgfscope}%
\pgfsetbuttcap%
\pgfsetroundjoin%
\definecolor{currentfill}{rgb}{0.000000,0.000000,0.000000}%
\pgfsetfillcolor{currentfill}%
\pgfsetlinewidth{0.803000pt}%
\definecolor{currentstroke}{rgb}{0.000000,0.000000,0.000000}%
\pgfsetstrokecolor{currentstroke}%
\pgfsetdash{}{0pt}%
\pgfsys@defobject{currentmarker}{\pgfqpoint{0.000000in}{-0.048611in}}{\pgfqpoint{0.000000in}{0.000000in}}{%
\pgfpathmoveto{\pgfqpoint{0.000000in}{0.000000in}}%
\pgfpathlineto{\pgfqpoint{0.000000in}{-0.048611in}}%
\pgfusepath{stroke,fill}%
}%
\begin{pgfscope}%
\pgfsys@transformshift{5.726543in}{0.580957in}%
\pgfsys@useobject{currentmarker}{}%
\end{pgfscope}%
\end{pgfscope}%
\begin{pgfscope}%
\definecolor{textcolor}{rgb}{0.000000,0.000000,0.000000}%
\pgfsetstrokecolor{textcolor}%
\pgfsetfillcolor{textcolor}%
\pgftext[x=5.726543in,y=0.483735in,,top]{\color{textcolor}\rmfamily\fontsize{10.000000}{12.000000}\selectfont \(\displaystyle {0.08}\)}%
\end{pgfscope}%
\begin{pgfscope}%
\definecolor{textcolor}{rgb}{0.000000,0.000000,0.000000}%
\pgfsetstrokecolor{textcolor}%
\pgfsetfillcolor{textcolor}%
\pgftext[x=3.124271in,y=0.304723in,,top]{\color{textcolor}\rmfamily\fontsize{10.000000}{12.000000}\selectfont Frequency (kyr\(\displaystyle ^{-1}\))}%
\end{pgfscope}%
\begin{pgfscope}%
\definecolor{textcolor}{rgb}{0.000000,0.000000,0.000000}%
\pgfsetstrokecolor{textcolor}%
\pgfsetfillcolor{textcolor}%
\pgftext[x=0.466444in,y=1.013548in,,bottom,rotate=90.000000]{\color{textcolor}\rmfamily\fontsize{10.000000}{12.000000}\selectfont \(\displaystyle I_\mathrm{Data}\) Power}%
\end{pgfscope}%
\begin{pgfscope}%
\pgfpathrectangle{\pgfqpoint{0.522000in}{0.580957in}}{\pgfqpoint{5.204543in}{0.865182in}}%
\pgfusepath{clip}%
\pgfsetrectcap%
\pgfsetroundjoin%
\pgfsetlinewidth{3.011250pt}%
\definecolor{currentstroke}{rgb}{1.000000,0.498039,0.054902}%
\pgfsetstrokecolor{currentstroke}%
\pgfsetdash{}{0pt}%
\pgfpathmoveto{\pgfqpoint{0.522000in}{0.628277in}}%
\pgfpathlineto{\pgfqpoint{0.603320in}{0.646939in}}%
\pgfpathlineto{\pgfqpoint{0.684640in}{0.632490in}}%
\pgfpathlineto{\pgfqpoint{0.765960in}{0.682177in}}%
\pgfpathlineto{\pgfqpoint{0.847280in}{0.669987in}}%
\pgfpathlineto{\pgfqpoint{0.928600in}{0.682920in}}%
\pgfpathlineto{\pgfqpoint{1.009920in}{0.627392in}}%
\pgfpathlineto{\pgfqpoint{1.091240in}{0.663109in}}%
\pgfpathlineto{\pgfqpoint{1.172560in}{1.406813in}}%
\pgfpathlineto{\pgfqpoint{1.253880in}{1.210572in}}%
\pgfpathlineto{\pgfqpoint{1.335200in}{0.745825in}}%
\pgfpathlineto{\pgfqpoint{1.416520in}{0.730686in}}%
\pgfpathlineto{\pgfqpoint{1.497840in}{0.671331in}}%
\pgfpathlineto{\pgfqpoint{1.579160in}{0.625561in}}%
\pgfpathlineto{\pgfqpoint{1.660480in}{0.623740in}}%
\pgfpathlineto{\pgfqpoint{1.741799in}{0.623845in}}%
\pgfpathlineto{\pgfqpoint{1.823119in}{0.649590in}}%
\pgfpathlineto{\pgfqpoint{1.904439in}{0.649185in}}%
\pgfpathlineto{\pgfqpoint{1.985759in}{0.631971in}}%
\pgfpathlineto{\pgfqpoint{2.067079in}{0.785188in}}%
\pgfpathlineto{\pgfqpoint{2.148399in}{0.820080in}}%
\pgfpathlineto{\pgfqpoint{2.229719in}{0.623816in}}%
\pgfpathlineto{\pgfqpoint{2.311039in}{0.629029in}}%
\pgfpathlineto{\pgfqpoint{2.392359in}{0.621435in}}%
\pgfpathlineto{\pgfqpoint{2.473679in}{0.621267in}}%
\pgfpathlineto{\pgfqpoint{2.554999in}{0.627046in}}%
\pgfpathlineto{\pgfqpoint{2.636319in}{0.633596in}}%
\pgfpathlineto{\pgfqpoint{2.717639in}{0.625926in}}%
\pgfpathlineto{\pgfqpoint{2.798959in}{0.626328in}}%
\pgfpathlineto{\pgfqpoint{2.880279in}{0.620308in}}%
\pgfpathlineto{\pgfqpoint{2.961599in}{0.622098in}}%
\pgfpathlineto{\pgfqpoint{3.042919in}{0.621697in}}%
\pgfpathlineto{\pgfqpoint{3.124239in}{0.620544in}}%
\pgfpathlineto{\pgfqpoint{3.205559in}{0.626976in}}%
\pgfpathlineto{\pgfqpoint{3.286879in}{0.641387in}}%
\pgfpathlineto{\pgfqpoint{3.368199in}{0.639592in}}%
\pgfpathlineto{\pgfqpoint{3.449519in}{0.661178in}}%
\pgfpathlineto{\pgfqpoint{3.530839in}{0.625083in}}%
\pgfpathlineto{\pgfqpoint{3.612159in}{0.620853in}}%
\pgfpathlineto{\pgfqpoint{3.693479in}{0.620797in}}%
\pgfpathlineto{\pgfqpoint{3.774799in}{0.621537in}}%
\pgfpathlineto{\pgfqpoint{3.856119in}{0.620406in}}%
\pgfpathlineto{\pgfqpoint{3.937439in}{0.626309in}}%
\pgfpathlineto{\pgfqpoint{4.018759in}{0.622874in}}%
\pgfpathlineto{\pgfqpoint{4.100078in}{0.620326in}}%
\pgfpathlineto{\pgfqpoint{4.181398in}{0.620443in}}%
\pgfpathlineto{\pgfqpoint{4.262718in}{0.621468in}}%
\pgfpathlineto{\pgfqpoint{4.344038in}{0.622356in}}%
\pgfpathlineto{\pgfqpoint{4.425358in}{0.620910in}}%
\pgfpathlineto{\pgfqpoint{4.506678in}{0.620733in}}%
\pgfpathlineto{\pgfqpoint{4.587998in}{0.620692in}}%
\pgfpathlineto{\pgfqpoint{4.669318in}{0.620796in}}%
\pgfpathlineto{\pgfqpoint{4.750638in}{0.623134in}}%
\pgfpathlineto{\pgfqpoint{4.831958in}{0.622218in}}%
\pgfpathlineto{\pgfqpoint{4.913278in}{0.621102in}}%
\pgfpathlineto{\pgfqpoint{4.994598in}{0.620416in}}%
\pgfpathlineto{\pgfqpoint{5.075918in}{0.620389in}}%
\pgfpathlineto{\pgfqpoint{5.157238in}{0.620743in}}%
\pgfpathlineto{\pgfqpoint{5.238558in}{0.620654in}}%
\pgfpathlineto{\pgfqpoint{5.319878in}{0.620356in}}%
\pgfpathlineto{\pgfqpoint{5.401198in}{0.620357in}}%
\pgfpathlineto{\pgfqpoint{5.482518in}{0.621046in}}%
\pgfpathlineto{\pgfqpoint{5.563838in}{0.622343in}}%
\pgfpathlineto{\pgfqpoint{5.645158in}{0.620652in}}%
\pgfpathlineto{\pgfqpoint{5.726478in}{0.620314in}}%
\pgfpathlineto{\pgfqpoint{5.736543in}{0.620328in}}%
\pgfusepath{stroke}%
\end{pgfscope}%
\begin{pgfscope}%
\pgfsetroundcap%
\pgfsetroundjoin%
\pgfsetlinewidth{1.003750pt}%
\definecolor{currentstroke}{rgb}{0.000000,0.000000,0.000000}%
\pgfsetstrokecolor{currentstroke}%
\pgfsetdash{{3.700000pt}{1.600000pt}}{0.000000pt}%
\pgfpathmoveto{\pgfqpoint{2.148399in}{2.629321in}}%
\pgfpathquadraticcurveto{\pgfqpoint{2.148399in}{1.605139in}}{\pgfqpoint{2.148399in}{0.580957in}}%
\pgfusepath{stroke}%
\end{pgfscope}%
\begin{pgfscope}%
\pgfsetroundcap%
\pgfsetroundjoin%
\pgfsetlinewidth{1.003750pt}%
\definecolor{currentstroke}{rgb}{0.000000,0.000000,0.000000}%
\pgfsetstrokecolor{currentstroke}%
\pgfsetdash{{3.700000pt}{1.600000pt}}{0.000000pt}%
\pgfpathmoveto{\pgfqpoint{3.286879in}{2.629321in}}%
\pgfpathquadraticcurveto{\pgfqpoint{3.286879in}{1.605139in}}{\pgfqpoint{3.286879in}{0.580957in}}%
\pgfusepath{stroke}%
\end{pgfscope}%
\begin{pgfscope}%
\pgfsetroundcap%
\pgfsetroundjoin%
\pgfsetlinewidth{1.003750pt}%
\definecolor{currentstroke}{rgb}{0.000000,0.000000,0.000000}%
\pgfsetstrokecolor{currentstroke}%
\pgfsetdash{{3.700000pt}{1.600000pt}}{0.000000pt}%
\pgfpathmoveto{\pgfqpoint{3.449519in}{2.629321in}}%
\pgfpathquadraticcurveto{\pgfqpoint{3.449519in}{1.605139in}}{\pgfqpoint{3.449519in}{0.580957in}}%
\pgfusepath{stroke}%
\end{pgfscope}%
\begin{pgfscope}%
\pgfsetroundcap%
\pgfsetroundjoin%
\pgfsetlinewidth{1.003750pt}%
\definecolor{currentstroke}{rgb}{0.000000,0.000000,0.000000}%
\pgfsetstrokecolor{currentstroke}%
\pgfsetdash{{3.700000pt}{1.600000pt}}{0.000000pt}%
\pgfpathmoveto{\pgfqpoint{3.937439in}{2.629321in}}%
\pgfpathquadraticcurveto{\pgfqpoint{3.937439in}{1.605139in}}{\pgfqpoint{3.937439in}{0.580957in}}%
\pgfusepath{stroke}%
\end{pgfscope}%
\begin{pgfscope}%
\pgfsetroundcap%
\pgfsetroundjoin%
\pgfsetlinewidth{1.003750pt}%
\definecolor{currentstroke}{rgb}{0.000000,0.000000,0.000000}%
\pgfsetstrokecolor{currentstroke}%
\pgfsetdash{{3.700000pt}{1.600000pt}}{0.000000pt}%
\pgfpathmoveto{\pgfqpoint{1.172560in}{2.629321in}}%
\pgfpathquadraticcurveto{\pgfqpoint{1.172560in}{1.605139in}}{\pgfqpoint{1.172560in}{0.580957in}}%
\pgfusepath{stroke}%
\end{pgfscope}%
\begin{pgfscope}%
\pgfsetrectcap%
\pgfsetmiterjoin%
\pgfsetlinewidth{0.803000pt}%
\definecolor{currentstroke}{rgb}{0.000000,0.000000,0.000000}%
\pgfsetstrokecolor{currentstroke}%
\pgfsetdash{}{0pt}%
\pgfpathmoveto{\pgfqpoint{0.522000in}{0.580957in}}%
\pgfpathlineto{\pgfqpoint{0.522000in}{1.446139in}}%
\pgfusepath{stroke}%
\end{pgfscope}%
\begin{pgfscope}%
\pgfsetrectcap%
\pgfsetmiterjoin%
\pgfsetlinewidth{0.803000pt}%
\definecolor{currentstroke}{rgb}{0.000000,0.000000,0.000000}%
\pgfsetstrokecolor{currentstroke}%
\pgfsetdash{}{0pt}%
\pgfpathmoveto{\pgfqpoint{5.726543in}{0.580957in}}%
\pgfpathlineto{\pgfqpoint{5.726543in}{1.446139in}}%
\pgfusepath{stroke}%
\end{pgfscope}%
\begin{pgfscope}%
\pgfsetrectcap%
\pgfsetmiterjoin%
\pgfsetlinewidth{0.803000pt}%
\definecolor{currentstroke}{rgb}{0.000000,0.000000,0.000000}%
\pgfsetstrokecolor{currentstroke}%
\pgfsetdash{}{0pt}%
\pgfpathmoveto{\pgfqpoint{0.522000in}{0.580957in}}%
\pgfpathlineto{\pgfqpoint{5.726543in}{0.580957in}}%
\pgfusepath{stroke}%
\end{pgfscope}%
\begin{pgfscope}%
\pgfsetrectcap%
\pgfsetmiterjoin%
\pgfsetlinewidth{0.803000pt}%
\definecolor{currentstroke}{rgb}{0.000000,0.000000,0.000000}%
\pgfsetstrokecolor{currentstroke}%
\pgfsetdash{}{0pt}%
\pgfpathmoveto{\pgfqpoint{0.522000in}{1.446139in}}%
\pgfpathlineto{\pgfqpoint{5.726543in}{1.446139in}}%
\pgfusepath{stroke}%
\end{pgfscope}%
\begin{pgfscope}%
\definecolor{textcolor}{rgb}{0.000000,0.000000,0.000000}%
\pgfsetstrokecolor{textcolor}%
\pgfsetfillcolor{textcolor}%
\pgftext[x=0.105637in,y=1.584568in,left,top]{\color{textcolor}\rmfamily\fontsize{11.000000}{13.200000}\bfseries\selectfont B}%
\end{pgfscope}%
\end{pgfpicture}%
\makeatother%
\endgroup%